\def \beq {\begin{equation}}
\def \bea {\begin{eqnarray}}
\def \eea {\end{eqnarray}}
\def \eeq {\end{equation}}

\newcommand{\ns}{\!\!\!}
\newcommand{\nsa}{\!\!\!\!}
\newcommand{\nse}{\ns&=&\ns}
\newcommand{\nn}{\nonumber}
\newcommand{\dc}{  \!\! \left \{
 \begin{array}{c} \scriptstyle  \\[-0.8cm] \scriptstyle
 \nsa MSM \nsa\\[-0.2cm]
                      \scriptstyle\nsa \nsa ECL \nsa \end{array}
 \right \}}
\newcommand{\ndc}{  \!\! \left \{
 \begin{array}{c} \scriptstyle  \\[-0.8cm] \scriptstyle
 \nsa MSM \nsa\\[-0.2cm]
                      \scriptstyle\nsa \nsa  -  \nsa \end{array}
 \right \}}
\documentstyle[12pt]{article}

\begin{document}
%
% >>>>>>>>>>>>>>>>>>>>>>>>>>>>>>>>>>>>>>>>>>>>>>>>>>>>>>>>>>>>>>>>>>>>
\begin{center}
%
%\hfill \hspace*{1cm}{\parbox[t]{2in}{
\hspace*{\fill}{\parbox[t]{1.4in}{
 DFPD 96/TH/18 \\
\hfill April 1996}}
\\[1cm]
{\large\bf THE EFFECTIVE CHIRAL LAGRANGIAN OF THE MSM
            AT ONE AND TWO LOOP ORDER   }
\\[1.3cm]
Joaquim Matias
\\[1.4mm]
 {\it Dipartimento di Fisica,
Universita  di Padova, Via F. Marzolo 8,
I-35131 Padova, Italy}\\
\end{center}

% >>>>>>>>>>>>>>>>>>>>>>>>>>>>>>>>>>>>>>>>>>>>>>>>>>>>>>>>>>>>>>>>>>>>
%
\vspace*{.7cm}
\centerline{\bf Abstract}
\vspace*{10mm}

The formalism of the matching conditions between transverse
connected Green functions is extended to include the two next to
leading corrections, namely the
 two-loop $M_{H}^{2}$ and the one-loop $1/M_{H}^{2}$ contributions
 to the
 coefficients
of the electroweak chiral lagrangian which are relevant to the LEP1
physics:
 $a_{0}$, ${\hat a}_{1}$ and
${\hat a}_{8}$.
We derive  general expressions for these three coefficients in terms
of just bare gauge boson self-energies. By means of the screening
theorem, it is shown that the same expressions  can be used
 to get directly from a MSM calculation,
the leading Higgs mass contribution to these coefficients at each
loop order.
In a more general framework, we solve the problems concerning the
loss of gauge invariance and the inclusion
of only gauge invariant operators by proposing a new formulation of
the
 matching conditions at two and higher loop order.
As an example of the usefulness
of using an electroweak chiral lagrangian to parametrize the MSM,
we will give a simple proof of an extra screening
 for the renormalized photon self-energy  in the on-shell
scheme at all orders. In addition it is shown the automatic
cancellation of the unphysical $M_{H}^4$
terms  in the other gauge boson self-energies at two-loops in this
scheme. Finally we will apply the obtained electroweak chiral
lagrangian
 to compute the different Higgs mass
contributions to the bosonic part of $\Delta \rho$, $\Delta r$ and
$\Delta \kappa$, analyzing carefully the
hierarchy between corrections.

\vspace{.5cm}
\hspace*{-1cm}{\rule{10cm}{.1mm}} \\
\vspace{.15cm}
E-mail address: matias@padova.infn.it

\vspace{1.2cm}
\pagebreak
\setcounter{page}{1}

\noindent { \Large \bf   Introduction}
\bigskip

\hspace{5mm} The electroweak theory has been tested at LEP1
with a high degree of accuracy. Instead, LEP1.5
and later on  LEP2 will search for new physics.
Still the precise mechanism of spontaneous symmetry breaking
remains one of the most interesting and hidden parts of that theory,
at least from an experimental point of view. The very few known
experimental facts, namely
the discovery of the massive $W^{\pm}$ and $Z$ gauge bosons
whose longitudinal parts can be interpreted as the Goldstone bosons
associated with the breaking of the global symmetry, the smallness
of the $\Delta \rho$ parameter, protected by the custodial
symmetry \cite{WeinbSuss} and the failure in finding a Higgs,
supersymmetric or technicolor particle up to a certain energy, leaves
the door open to all sort of speculations on possible mechanisms
compatibles
 with them.

For that reason it
 can be useful to find some general tool able to perform a `natural'
splitting between the `known physics' (where `known' means
experimentally tested) and possible effects,
 in our present available energies, coming from
 `new physics' in this experimental sense (including the case of a
Higgs of the Minimal Standard Model -MSM-).
 We have such a tool at hand and it
is called Chiral Perturbation Theory \cite{GL,TP}, or more
specifically, the
electroweak chiral lagrangian (see \cite{Fe} for a nice review on the
topic).

In such lagrangian the splitting between `known' and `new' physics
is
made explicit. The known spectra of particles and symmetries are the
basic building blocks whereas the information concerning the
particular breaking mechanism will fit into the coefficients of the
operators compatibles with the symmetries (which we will usually
refer as chiral coefficients).
These coefficients will
collect
the `low energy effects'
 coming from still undiscovered new heavy degrees of
freedom (Higgs in the MSM, techniparticles in
technicolor theories \cite{Tec,Pesk}, $Z^{\prime}$ in $E_{6}$ models
\cite{Zpri}, $W^{\prime}$ in leftright models \cite{lr}, $\ldots$).

We will concentrate in getting these non-decoupling  effects
\cite{AC} of the heavy degrees of freedom in one
of the possible models: the large Higgs mass limit of the Minimal
Standard Model.

In an early work Appelquist and Bernard \cite{Appelq} for an SU(2)
Yang-Mills theory and later on Longhitano \cite{Long} for
the complete $SU(2)_{L} \times U(1)_{Y}$ gauge group found the
leading one-loop logarithmic dependence on the Higgs mass of
the chiral coefficients. Later on
 the finite piece accompanying this leading ${\rm Log}$ was
obtained by using functional methods in \cite{GL}. More recently in a
series of papers \cite{MJ1,MJ2,EM,TOT} a different technique, so
called matching conditions has been used to derive both one-loop
contributions.

The purpose of this paper will be twofold. On the one hand,
we will extend and apply the technique
of the matching conditions to obtain for the first
time
the two first subleading corrections,
namely the one-loop
$1/M_{H}^{2}$ contribution and the two-loop $M_{H}^{2}$,
 entering the three chiral
coefficients relevant to LEP physics.
On the other hand, it will be shown how
one can benefit from the `natural' splitting explained above
 in getting information
on the  Higgs mass
contribution to the self-energies.

Some results concerning the three point functions are also given.

 We will pay
special attention to the subtleties concerning gauge invariance in
the
matching conditions at two-loops proposing a new formulation of these
equations. In addition we will comment on the scheme independence of
the chiral coefficients and the hierarchy between the
different subleading Higgs corrections.

 We will consider a scenario with a Higgs `heavy' enough
to fulfill the necessary requirement for having a chiral lagrangian,
i.e.,
the existence of a mass gap between the heavy and light degrees of
 freedom, but sufficiently `light'  not to break the perturbative
expansion.
Moreover due to the strong suppression of the
$M_{H}^{2}$ contributions coming from two loops the range of energies
where a chiral description can be applied is very large.

The paper is organized in the following way. We start
in section 1 presenting the
two theories that we want to match. Then we will make some general
comments on the matching conditions in section 2.
We will show in  section 3  the matching equations at one-loop and
we will give an expression for the  coefficients of the chiral
operators in terms of just bare self-energies.
 We will analyze the {\bf complete} Higgs one-loop
contribution, pointing out the gauge invariance of the transverse
part.
Afterwards we will crumble the different Higgs contributions entering
into the chiral coefficients, mainly the leading ${\rm Log} M_H^2$
and the first
$1/M_H^{2}$.
We will end this section with a brief discussion on the
dimension six
operators. In section 4 we will address the problem of how
to write
the matching conditions at two loops.
We will give a simple solution and afterwards we will show how the
$M_H^2$ contribution enters into the chiral coefficients.
 Then we will use the obtained chiral lagrangian to prove an extra
 screening
of the renormalized photon self-energy in the on-shell scheme
to all orders.
 We will discuss in section 5 on the hierarchy of corrections,
i.e., the competition between the two-loop
$M_H^2$ contribution and
the one-loop $1/M_H^2$ correction.
 We will  analyze their relative importance
in relation with the quantities
 $\Delta \rho$, $\Delta r$ and $\Delta \kappa$
 as a function of the Higgs mass. We will finally
jump into the conclusions.

In Appendix A the complete set of $SU(2)_{L}\times U(1)_{Y}$
gauge invariant operators of the electroweak chiral lagrangian (up to
order $p^{4}$) is given.
And the details of the large Higgs mass expansion of the one-loop
 gauge self-energies are explained
in Appendix B. \bigskip

\noindent { \Large \bf 1  MSM and electroweak
chiral lagrangian}
\bigskip

We shall start our discussion by presenting the two theories
that we want to link, the
Minimal Standard Model as the fundamental one and the electroweak
chiral lagrangian as its effective low energy realization.

The lagrangian of the MSM can be written in a
general
way, without specifying the particular realization chosen for the
scalar particles \cite{Long}
\bea
\label{eq01}
&{\cal L}_{MSM}=\ns&
  +{1 \over 4} {\rm Tr}D_{\mu} M^{\dag} D^{\mu} M - {1 \over 4}
{\lambda}{( {1 \over 2} {\rm Tr} M^{\dag} M  + {\mu^2 \over \lambda}
)}^{2} \nn \\ &&-{1 \over 2} {\rm Tr} W_{\mu\nu} W^{\mu\nu}
-{1 \over 4} B_{\mu\nu} B^{\mu\nu}
+{\cal L}_{GF}
+ {\cal L}_{FP}, \eea
where $\lambda$ stands for the Higgs self-coupling and $\mu^{2}$ is
related to the vacuum expectation value by
$v=\sqrt{-\mu^{2}/\lambda}$.
The matrix $M$  collects all scalar fields and the covariant
derivative acting over this matrix is defined by
$D_{\mu}M=\partial_{\mu}M+i g W_{\mu} M - i g^{\prime} B_{\mu} M
{1 \over 2}\tau^{3}$ with $B_{\mu}$ and  $W_{\mu}={1 \over
2}W_{\mu}^{i}
\tau^{i}$ the vector boson fields. $\tau^{i}$ are the Pauli matrices.

In choosing one or another
parametrization for this $2 \times 2$ matrix $M$ one is changing the
realization of the
scalar fields. In between all possible parametrizations we will
comment
 on two
\beq  \label{eq02}
M_{1}=\sigma + i {\vec \omega} {\vec \tau} \quad \quad \quad
M_{2}=\rho \, U = \rho \, {\rm  exp}\, (i \, {\vec \pi} {\vec \tau} /
v), \eeq
where $\sigma$ and $\rho$ represent the Higgs field while ${\vec
\omega}$ and ${\vec \pi}$ stand for the Goldstone boson fields.
The first one, is a linear realization. If one uses this
parametrization in
(\ref{eq01}) the linear presentation of the MSM is recovered (see for
instance \cite{Muta}).
On the contrary, the second one gives a
nonlinear realization for the Goldstone bosons.
We will choose
this second  parametrization $M_{2}$ for the
scalar sector of the Standard Model.
Notice that in the chiral lagrangian the Goldstone bosons transforms
 nonlinearly too.

Along this paper it will be shown that this parametrization apart
  from making the connection between both theories
(MSM and its effective electroweak chiral lagrangian)
 more natural,
 it will be useful in pointing out some conceptual problems
that were previously overlooked. Moreover we will see that from a
technical point of view it has
an obvious advantage in front of the linear parametrization of the
MSM regarding the matching conditions.

In these variables the MSM lagrangian is written
\bea \label{eq03}
&{\cal L}_{MSM}=\ns&{1 \over 2} \partial_{\mu} \rho \partial^{\mu}
\rho -\rho {\lambda v} (v^2+{\mu^2\over \lambda})
-{1 \over 2} \rho^2 (\mu^2 + 3 v^2 \lambda)
- \lambda v \rho^3 - {1\over 4}\lambda\rho^4 \nn \\
&&+{1 \over 4} (\rho+v)^2 {\rm Tr}D_{\mu} U^{\dag} D^{\mu} U
-{1 \over 2} {\rm Tr} W_{\mu\nu} W^{\mu\nu}
-{1 \over 4} B_{\mu\nu} B^{\mu\nu}
+
{\cal L}_{GF}
+ {\cal L}_{FP}, \qquad \eea
with $D_{\mu} U=\partial_{\mu} U + {1 \over 2} i g W_{\mu}^{i}
\tau^{i} U(x) - {1 \over 2} i g^\prime B_{\mu} U(x) \tau^{3}  $.
For the gauge-fixing  we will take the
 usual gauge fixing
term
 \bea \label{eq06}
{\cal L}_{GF}\ns &\equiv \ns&
 -{1\over {2}}\sum_{1=1,3}F^iF^i-{1\over {2}}F^0F^0 \nn \\
\nse -{1 \over 2 \xi^{W}}\sum_{i=1,3} {(\partial^{\mu}
W_{\mu}^{i}
- {1 \over 2} g v \xi^{W} \pi^{i})}^{2}
-{1 \over 2\xi^{B}}{(\partial^{\mu} B_{\mu} +
{1 \over {2 }} g^\prime v \xi^{B} \pi^{3})}^{2} ,
\eea
having in mind when constructing the Faddeev-Popov
 \bea \label{eq07}
&{\cal L}_{FP}=\ns& \sum_{\alpha , \beta=i,0} {\bar c}^{\alpha}
{\delta F^{\alpha} \over \delta \theta^{\beta}} c^{\beta},
   \eea
that the Goldstone bosons will transform  nonlinearly
\bea  \label{eq11}
\delta \pi^{i}&=& {1 \over 2} ( v (\theta^{i}-\theta^{0} \delta^{i3})
 - \epsilon^{ijk} \theta^{j} \pi^{k} + \theta^{0}(\pi^{2}\delta^{i1}-
\pi^{1}\delta^{i2})) \nn \\
&+& {1\over 6 v} \pi^{j} \pi^{l} (\theta^{k}(\delta^{li}\delta^{kj}-
\delta^{jl}\delta^{ki})-\theta^{0}(\delta^{j3}\delta^{li}-
\delta^{i3}\delta^{jl} )) + \ldots
\eea
where $\theta^{i}$ and $\theta^{0}$ are the  parameters of the
$SU(2)$ and $U(1)$ transformation respectively.
An expansion up to four fields of lagrangian (3)  can be found in
Appendix D of \cite{Eq}.

The electroweak chiral lagrangian is constructed by
relaxing the renormalizability constrain and by writing all possible
operators,
up to a certain order in momenta,
that are
consistent with the symmetries, i.e., Lorentz, C and P  and $SU(2)_L
\times U(1)_Y$ having  assumed a pattern of spontaneous symmetry
breaking from a global $SU(2)_L \times SU(2)_R$ group to a vectorial
$ SU(2)_V$ \cite{Long}.

 This lagrangian is, up to order $p^4$, given by
\beq
  \label{eq08}
{\cal L}^{eff}=
+{v^2 \over 4} {\rm Tr} D_{\mu} U^{\dag}
D^{\mu} U
-{1 \over 2} {\rm Tr} W_{\mu\nu} W^{\mu\nu}
-{1 \over 4} B_{\mu\nu} B^{\mu\nu}
+{\cal L}_{GF}+{\cal L}_{FP}+
\sum_{i=0,13} a_i {\cal L}_{i},
\eeq
where the last fourteen {\bf gauge invariant} operators
are listed in Appendix A.

 It can be proved \cite{Fe}
 that if one uses the equations of motion over the previous set of
operators  they
 reduce to just eleven operators (${\cal L}_{11}$ and ${\cal L}_{12}$
vanishes
identically and ${\cal L}_{13}$ can be written as a linear
combination
of the ${\cal L}_{1} \ldots {\cal L}_{10}$  operators).  While
it is allowed to use the equations of motion at one-loop
 it is not at two loops. As we will be interested in performing a
two-loop computation,
 we will never use the equations of motion over the ${\cal O}(p^4)$
operators.

The renormalized lagrangians will be obtained by redefining the
fields and introducing the renormalization constants \cite{Long} in
both theories
 \begin{eqnarray} \label{eq09}
W_{\mu}^{i}&\rightarrow&Z_{W}^{1 \over 2} W_{\mu}^{i}   \nn   \\
B_{\mu}&\rightarrow&Z_{B}^{1 \over 2} B_{\mu}     \nn   \\
\pi^{i} &\rightarrow& Z_{\pi}^{1 \over 2} \pi^{i}         \nn  \\
g &\rightarrow & Z_{W}^{-{1 \over 2}} g \left( 1 - {\delta g \over g}
\right) \nn \\ g\prime &\rightarrow & Z_{B}^{-{1 \over 2}} g^{\prime}
\left( 1 - {\delta g^{\prime} \over g^{\prime}} \right) \nn \\
v &\rightarrow & Z_{\pi}^{1 \over 2} v \left( 1 - {\delta v \over v}
\right). \end{eqnarray}

In the electroweak chiral lagrangian, in addition, the chiral
coefficients require
 a further renormalization $a^{b}_{i}=a^{r}_{i}+\delta a_{i}$.
For a discussion on renormalization procedures in Chiral
Perturbation Theory see for instance \cite{RM}.
\bigskip

\noindent{\Large \bf 2  Matching Conditions and Gauge Invariance}
\bigskip

One of the possible ways, although not the only one (see
\cite{DG95} \cite{Nyf}), to connect a fundamental theory with its
low energy effective description are the
matching conditions. In a general framework they consist of
a set of equations where one imposes that some quantity evaluated
with the fundamental
theory should coincide with the same quantity obtained
using the electroweak chiral lagrangian (that will depend on the
$a$'s)
up to a certain order (in inverse powers of the heavy mass particle)
and up to a certain scale. Depending on the type of object we choose
to
match, i.e., the level at which to impose the matching one needs to
be
more careful. The care has to deal with the observable or non
observable nature of the objects that one uses for making the
matching.

There are two types of objects that one could think of using in the
matching. The first, of course, is a S-matrix element which is a
gauge invariant quantity.
 The other possibility,
which is the one we will take, is to
use as the matching object  connected Green functions with external
gauge fields.
In that case we will be able to
deal with off-shell quantities too.

There is, indeed, a third approach
using One-Particle-Irreducible functions\cite{MJ1,MJ2}
that
 gives the same results for the chiral coefficients as far as the
transverse part is concerned. However it leads to inconsistent
equations\cite{EM}, already at one-loop, for the longitudinal ones
under a simple change of variables as the one of eq.(\ref{eq02}).

Let us make some comments on the properties of the connected Green
functions concerning the matching conditions. On the one hand, this
type of Green
function exhibits the nice property of being  invariant
under the redefinition of
the scalar fields given by eq.(\ref{eq02}) (
this property  has been checked explicitly for the two and three
point
functions at one-loop in \cite{EM}). On the other hand, if one uses
an $R_{\xi}$  or some other type of gauge (except for the
background field gauge) to calculate
a Green function, it is clear that gauge invariance is lost and
one is dealing then with a BRST invariant quantity.
It implies automatically that if we use the
complete two point connected Green function
(transverse plus longitudinal) in the matching conditions
they turn out to be BRST invariant equations.
 This could be, in principal, a disadvantage compared to the
$S$ matrix elements (which define gauge invariant matching equations)
because
one is interested  in consider only gauge
invariant
operators $a$'s and not all the possible BRST operators. However,
there is a nice way to bypass this problem \cite{EM}
and it is to consider a subset of gauge invariant matching equations.
We will see from a direct computation that the contribution at
one-loop coming from the Higgs to the {\bf transverse part of the
connected Green functions}
is still gauge invariant. Then if one restricts the matching
conditions at one-loop to the transverse part it is consistent
to consider {\bf just} the gauge invariant operators
of Appendix A.

On the contrary if one includes in the matching the longitudinal
parts of the Green functions
one gets into troubles because the longitudinal parts  are affected
 by the particular choice  of the gauge in
the fundamental and in the effective theory. In the
case of using One-Particle-Irreducible functions it is even worst
because their longitudinal
parts  also depend on the particular parametrization used for the
scalars.
 It is clear then that one cannot make absolute
statements (independently on the gauge, scheme, parametrization,
etc.) about the values of the coefficients entering
such parts, because one can fall easily into inconsistencies
regarding
 the matching
conditions \cite{EM}. On the other hand one would expect that the
coefficients entering the longitudinal parts ($a_{11},a_{12}$ and all
only BRST invariant operators that we will call
$b$ operators\footnote{Notice that we distinguish between operators
that are gauge invariant ($a$'s) from those that are only BRST
invariant ($b$'s). From now on wherever we talk about BRST or extra
BRST invariant operators we refer to the $b$ operators.})
will be irrelevant at the level of observables. So the safest
position
will be clearly not to consider the longitudinal parts (unless one
considers the complete set of BRST invariant operators).
In that way,
 at the end of the day, we have been
 self-consistent including in the restricted gauge invariant matching
equations only gauge invariant
operators and, in addition, we are not losing any physical
information.

\bigskip
\noindent{\Large \bf 3  One-loop matching conditions  }
\bigskip

Let's now see the explicit expression of the matching equations
between  the gauge field self-energies. From now on and according to
the previous discussion when we  refer to gauge field self-energies
we will always intend the transverse ones. In a compact notation the
matching conditions are simply
 \beq
\label{eq016}
{\hat \Sigma}^X_{MSM} (p^2;\mu)= {\hat \Sigma}^X_{ECL} (p^2;\mu)
  \qquad \qquad \mu \leq M_H ,
\eeq
where ${\hat \Sigma}^{X}_{MSM}$ and ${\hat \Sigma}^{X}_{ECL}$ stands
for the {\bf renormalized} self-energies of
gauge fields  ($X=WW,ZZ,\gamma \gamma,\gamma Z$) in the Minimal
Standard Model and the electroweak chiral lagrangian respectively.
The explicit expressions of the renormalized self-energies
at one-loop are
\begin{eqnarray}
\label{eq017}
{\hat
\Sigma}^{WW}_{\dc}
(p^{2};\mu^{2})&=&{\Sigma_{0}}^{WW}_{ \dc
      }(p^{2})
- {\delta M_{W}^2}_{\dc  } -\delta t_{\ndc} M_{W}^{2} \nn \\ & & +
{\delta Z_{W}}_{ \dc} (p^2 - M_{W}^{2}),   \nn \\
{\hat
\Sigma}^{ZZ}_{\dc  }(p^{2};\mu^{2})&=&{\Sigma_{0}}^{ZZ}_{\dc}(p^{2})
- {\delta
M_{Z}^2}_{\dc  } -\delta t_{\ndc} M_{Z}^{2} \nn \\
& & + ( c^2 {\delta Z_{W}}_{\dc  }   + s^2 {\delta
Z_{B}}_{\dc  }) (p^2 - M_{Z}^{2}), \nn  \\
{\hat
\Sigma}^{\gamma
\gamma}_{\dc}(p^{2};\mu^{2})&=&{\Sigma_{0}}^{\gamma
\gamma}_{\dc}(p^{2})
+ ( s^2 {\delta Z_{W}}_{\dc} + c^2 {\delta Z_{B}}_{\dc})
p^{2}, \nn
\\ {\hat   \Sigma}^{\gamma
Z}_{\dc}(p^{2};\mu^{2})&=&{\Sigma_{0}}^{\gamma
Z}_{\dc}(p^{2})+ {\frac{1}{4} g g^{\prime} v^{2}} ( {\frac{\delta g}{
g}} -
{\frac{\delta g^{\prime}}{ g^{\prime} }} )_{\dc} \nn \\ & & + s \, c
({\delta Z_{W}}_{\dc}  -{ \delta Z_{B}}_{\dc}) p^{2}, \end{eqnarray}
where $\delta t$ stands for the tadpole counterterm and $s,c$ are the
sinus and cosinus of the Weinberg angle respectively.
The mass renormalization constants can be written easily
in terms of the renormalization constants of eq.(\ref{eq09})
\bea
{\delta M_{W}^{2} \over M_{W}^{2}}&=&\delta Z_{\pi} - \delta Z_{W} -
2 {\delta v \over v} -2 {\delta g \over g},  \nn \\
{\delta M_{Z}^{2} \over M_{Z}^{2}}&=&\delta Z_{\pi} - c^{2} \delta
Z_{W} -s^{2} \delta Z_{B} - 2 {\delta
v \over v} -2 c^{2} {\delta g \over g} - 2 s^{2} {\delta g^{\prime}
\over g^{\prime}}.
\eea
The bare self-energies of the MSM can be split up into two pieces
\beq \label{eq018}
{\Sigma_{0}}_{\{ MSM\} }(p^{2})={\Sigma_{0}}_{(L) \{ MSM\}}(p^{2})
+{\Sigma_{0}}_{(H)\{ MSM\} }(p^{2}). \eeq
${\Sigma_{0}}_{(L) \{ MSM \} }$ collects all the diagrams with
only light
particles, whereas ${\Sigma_{0}}_{(H) \{ MSM \} }$ includes  the
rest of the diagrams  with at least one Higgs running inside the
loop.

In the chiral lagrangian one can similarly distinguish between two
contributions
\beq  \label{eq019}
{\Sigma_{0}}_{\{ ECL \} }(p^{2})={\Sigma_{0}}_{(L) \{ ECL \}}(p^{2})
+{\Sigma_{0}}_{(H=a)\{ ECL \} }(p^{2}), \eeq
with ${\Sigma_{0}}_{(L) \{ ECL \} }$ representing as in the MSM
the
diagrams with only light particles and ${\Sigma_{0}}_{(H=a)\{ ECL\}
}$ stands for the contribution of the fourteen operators listed
in Appendix A, that collect the non-decoupling effects of the Higgs
in the chiral lagrangian.

It is remarkable to notice by a direct inspection of lagrangian
(\ref{eq03}) and lagrangian (\ref{eq08}) that if we choose the same
gauge-fixing eq.(\ref{eq06}) and Faddeev-Popov term eq.(\ref{eq07})
in both lagrangians the following equality holds
\beq \label{eq020}
{\Sigma_{0}}_{(L) \{ MSM \} }(p^{2})={\Sigma_{0}}_{(L) \{ ECL \}
}(p^{2}). \eeq
The reason being that  in the nonlinear representation $M_{2}$ for
the Goldstone bosons of the MSM (\ref{eq03}) the Feynman rules
involving just
light fields become the same as in the electroweak chiral lagrangian.
This is not true if, instead, we use the linear
parametrization $M_{1}$.

The explicit contribution of the chiral operators to the bare  gauge
field self-energies is \cite{EM}
\bea
\label{eq021}
{\Sigma_{0}}^{WW}_{(H=a)\{ECL\}}(p^{2})&=&0, \nn \\
{\Sigma_{0}}^{ZZ}_{(H=a)\{ECL\}}(p^{2})&=&+2 M_{Z}^{2} a_{0} +
p^{2} (c^{2}
g^{2} a_{8} + 2 s^{2} g^{2} a_{1} + (g^{2} + g^{\prime 2})
a_{13}), \nn
\\ {\Sigma_{0}}^{\gamma \gamma}_{(H=a)\{ECL\}}(p^{2})&=& + p^{2}
s^{2} g^{2} (a_{8} - 2 a_{1}),   \nn
\\ {\Sigma_{0}}^{\gamma Z}_{(H=a)\{ECL\}}(p^{2})&=& + p^{2} ( s c
g^{2} a_{8} - (c^{2} - s^{2}) g g^{\prime} a_{1} ). \eea

In order to determine these coefficients
we will  impose the matching equations ({\ref{eq016}}) using
eq. ({\ref{eq017}}) and taking into account, on the one hand,
that the renormalization constants in both theories are, in
general, different ($\delta
Z_{\{MSM\} } \neq \delta Z_{\{ ECL \} }$) and, on the other, the
exact cancellation
of the light bare self-energies of both theories (\ref{eq020})
in the nonlinear representation of the MSM (\ref{eq03}).

If one solves these set of equations for the two-point functions, one
ends up with the following expressions for  the $a$'s in
terms of just the heavy part of the bare self-energies of the MSM
($\Sigma^{X}_{0 (H) \{MSM\}}$)
\bea  \label{eq025}
a_{0}&\ns=\ns&-{1 \over 2 M_{Z}^{2} c^{2}} \biggl(
{\Sigma_{0 (H) }^{WW}}
- c^{2}  {\Sigma_{0 (H)}^{ZZ}} - 2 s c
{\Sigma_{0 (H) }^{\gamma Z}} \biggr) \bigg\vert_{p^{2}=0},
\nn \\
{\hat a}_{1}&\ns=\ns&+{1 \over g^{2}} \left( c^{2} {\partial
{\Sigma_{0 (H)}^{ZZ}}
\over \partial p^{2}}  - c^{2} {\partial
{\Sigma_{0 (H)}^{\gamma \gamma}}
\over \partial p^{2}} - {c \over s} (c^{2}-s^{2}){\partial
{\Sigma_{0 (H)}^{\gamma Z}} \over \partial p^{2}} \right)
\bigg\vert_{p^{2}=0},
\nn \\
{\hat a}_{8}&\ns=\ns&+{1 \over g^{2}} \left( c^{2} {\partial
{\Sigma_{0 (H)}^{ZZ}} \over \partial p^{2}}  + s^{2}
{\partial
 {\Sigma_{0 (H)}^{\gamma \gamma}}
\over \partial p^{2}} + {2 s c} {\partial
{\Sigma_{0 (H)}^{\gamma Z}} \over \partial p^{2}}-
{\partial
{\Sigma_{0 (H)}^{WW}} \over \partial p^{2}} \right)
\bigg\vert_{p^{2}=0},
 \eea %
where the suffix ${\{MSM\}}$ has been omitted to keep the
expression manageable. ${\hat a}_{1}$ and ${\hat a}_{8}$ stands for
the combinations $a_{1}+a_{13}$ and $a_{8}+a_{13}$ respectively that
enter the observables always in this precise way. The previous
equations have the remarkable property of making explicit at one-loop
the independence on the scheme
chosen to renormalize the fundamental
and the effective theory of the $a$ coefficients.

The same procedure explained above for the two-point functions
can
be followed exactly to get the heavy particle contribution to the
$a$'s
 entering the
three point functions. There the matching equations would read
 \beq
\label{eq027}
{\hat \Gamma}^Y_{\lambda \mu \nu (T)\{ MSM\}}
(p_{1},p_{2},p_{3};\mu)=
{\hat \Gamma}^Y_{\lambda \mu \nu (T) \{ECL\}} (p_{1},p_{2},p_{3};\mu)
  \qquad \qquad \mu \leq M_{H},
\eeq
where $Y$ stands for the vertex $ZW^{+}W^{-}$ and $\gamma W^{+}W^{-}$
and $p_{1}$, $p_{2}$ and $p_{3}$ are the momenta of the three
incoming
particles ($Z/\gamma$,$W^{+}$ and $W^{-}$ respectively). The subindex
(T) means that one should restrict
to the non-vanishing
structures when the gauge condition $\epsilon^{\mu}(p^{i}) p^{i}_{
\mu}=0$ is applied.
This is the way of restricting to a gauge invariant subset of
equations for the three point functions.
Following
the same steps as before for the three point functions one would get
\bea \label{eq028}
a_{2}&=& -{c \over g^{3}} \left( - {c \over s} C_{2}^{\gamma W^{+}
W^{-}} + C_{2}^{Z W^{+} W^{-}}  - {c \over s} C_{1}^{\gamma W^{+}
W^{-}} + C_{1}^{Z W^{+} W^{-}} \right) + {\hat a}_{1},  \nn \\
a_{3}&=& -{c \over g^{3}} \left( -{c \over s} C_{1}^{\gamma W^{+}
W^{-}}+C_{1}^{Z W^{+} W^{-}} \right), \nn \\
a_{9}&=& -{1 \over g^{3}} \left( s C_{2}^{\gamma W^{+}
W^{-}}+c C_{2}^{Z W^{+} W^{-}} \right) + {\hat a}_{8},
\eea
with
\bea  \label{eq029}
C_{1}^{Y}={1 \over 18} {{p_{1}^{\nu}} \over p_{3}^2 - p_{2}^{2}}
\left( \Gamma_{0 (H)\mu \mu \nu}^{Y} + \Gamma_{0 (H)\mu \nu \mu}^{Y}
- 5 \Gamma_{0 (H)\nu \mu \mu}^{Y} \right), \nn \\
C_{2}^{Y}={1 \over 6} {p_{1}^{\nu}} \left(
{1 \over p_{1}^{2}} ( \Gamma_{0 (H)\mu \nu \mu}^{Y} - \Gamma_{0
(H)\mu \mu
\nu}^{Y} ) + {3 \over p_{3}^{2} - p_{2}^{2}} \Gamma_{0 (H)\nu \mu
\mu}^{Y} \right). \eea
$\Gamma_{0 (H)}$ stands for the contribution coming from the diagrams
with at least one Higgs inside. Notice that the subscript (T) has
disappeared in the $\Gamma$'s because
in these equations the longitudinal part drops out automatically.
${\hat a}_{1}$ and ${\hat a}_{8}$ are given by equations
(\ref{eq025}).

\bigskip
\noindent{\large \bf 3.1 Leading one-loop $M_{H}$ contribution}
\bigskip

 This general formalism can be applied to get the Higgs
contribution to the $a$'s.
We will restrict from now on to the two-point functions
 to be able to develop in detail an explicit example,
however
the same technique can be applied to the three and four point
functions.

The first step will be to get the Higgs contribution to the four
self-energies with external gauge fields.
 It is easy to see that while the photon and mixed $Z$-photon
self-energies do not have any Higgs contribution at 1-loop order, the
$W$ and $Z$ self-energies get contributions from the three diagrams
drawn in Fig.1.

Their explicit expression
in an arbitrary $R_{\xi}$ gauge are\footnote{Similar expressions in
t'Hooft-Feynman gauge but using the Passarino-Veltman $B_{i}$
functions can be found in \cite{PV}.},
 \bea \label{eq030}
 {\Sigma_{0 \, (H) \{ MSM \}}^{WW}}
(p^{2})&=&  {g^{2} \over 16 \pi^2}
\left\{ C_{UV} M_{W}^{2} - {1 \over 4} ( C_{UV} + 1 ) (M_{W}^{2} - {
p^{2} \over 3} ) \right. \nn \\
& & \left. - {1 \over 4} M_{H}^{2}{\rm Log} {M_{H}^{2} \over \mu^{2}}
+ \int_{0}^{1} dx \left[ {1 \over 2} D_{2}(M_{W}^{2}) - M_{W}^{2}
\right] {\rm Log} {D_{2}(M_{W}^{2})\over \mu^{2}} \right\}, \nn \\
 {\Sigma_{0 \, (H) \{ MSM \}}^{ZZ}}
(p^{2})&=&  {g^{2} \over 16 \pi^2
c^{2}}
\left\{ C_{UV} M_{Z}^{2} - {1 \over 4} ( C_{UV} + 1 ) (M_{Z}^{2} - {
p^{2} \over 3} ) \right. \nn \\
& & \left. - {1 \over 4} M_{H}^{2} {\rm Log} {M_{H}^{2} \over
\mu^{2}}
+ \int_{0}^{1} dx \left[ {1 \over 2} D_{2}(M_{Z}^{2}) - M_{Z}^{2}
\right] {\rm Log} {D_{2}(M_{Z}^{2}) \over \mu^{2}} \right\}, \nn \\
 {\Sigma_{0 \, (H) \{ MSM \}}^{\gamma \gamma}}
(p^{2})&=& 0, \nn \\
 {\Sigma_{0 \, (H) \{ MSM \}}^{\gamma Z}}
(p^{2})&=& 0,
 \eea
where
\beq \label{eq31}
D_{2}(M^{2})=M_{H}^{2} + (M^{2} - M_{H}^{2} - p^{2}) x + p^{2}
x^{2} \eeq
and $C_{UV}={1 / \epsilon} - \gamma + {\rm Log} 4 \pi$.
Notice that no tadpole  is included in the list of diagrams of
Fig.1 the reason being that they are exactly canceled  by the
tadpole counterterm  $\delta t$. Moreover, the only
place where $\delta t$ would enter is through the $W$ and $Z$
self-energies at $p^{2}=0$, and both contributions would cancel each
other into the $a_{0}$ coefficient anyway.

It is
important to point out that  eq.(\ref{eq030})
although
calculated in a general $\xi$ gauge happen to be $\xi$ independent to
all orders in the $M_{H}$ expansion.
This result implies the gauge independence of the matching conditions
between the transverse gauge field self-energies at one-loop to all
orders in the $1/M_{H}^{2}$ expansion.

If we now expand the self-energies given by eq.(\ref{eq030}) up to
the leading $M_{H}$ correction by using the formulae of Appendix B
\bea \label{eq031}
&
 {\Sigma_{0 \, (H) \{ MSM \}}^{WW (1)}}
(p^{2})=\ns&
- {g^2 \over 16 \pi^2} \left\{ {M_{H}^{2} \over 8} - {3 \over 4}
M_{W}^{2}
\left( C_{UV} - {\rm Log} {M_{H}^{2} \over \mu^{2}} + {5 \over 6}
\right) \right. \nn \\ &&- \left. {1 \over 12}
p^{2} \left( C_{UV} - {\rm Log} {M_{H}^{2} \over \mu^{2}} + {5 \over
6} \right) \right\},
 \nn \\
&
 {\Sigma_{0 \, (H) \{ MSM \}}^{ZZ (1)}}
(p^{2})=\ns&
- {g^2 \over 16 \pi^2 c^{2}} \left\{ {M_{H}^{2} \over 8} - {3 \over
4} M_{Z}^{2}
\left( C_{UV} - {\rm Log} {M_{H}^{2} \over \mu^{2}} + {5 \over 6}
\right) \right. \nn \\ &&- \left. {1 \over 12}
p^{2} \left( C_{UV} - {\rm Log} {M_{H}^{2} \over \mu^{2}} + {5 \over
6} \right) \right\},
 \nn \\
&
{\Sigma_{0 \, (H) \{ MSM \}}^{\gamma \gamma (1)}}
(p^{2})=\ns& 0, \nn \\
&
 {\Sigma_{0 \, (H) \{ MSM \}}^{\gamma Z (1)}}
(p^{2})=\ns& 0,
\eea
one obtains after substituting eq.(\ref{eq031}) into
 eq. (\ref{eq025})
\bea \label{eq032}
a_{0}^{b \, (1)}(\mu)&=& {g^{\prime 2} \over 16 \pi^{2}}{3 \over 8}
\left( C_{UV} -{\rm Log} {M_{H}^{2}\over \mu^{2} } + {5 \over 6}
\right), \nn
\\ {\hat a}_{1}^{b \, (1)}(\mu)&=& {1 \over 16 \pi^{2}}
{1 \over 12} \left(
C_{UV}-{\rm Log} {M_{H}^{2}\over \mu^{2}} + {5 \over 6} \right), \nn
\\
{\hat a}_{8}^{b \, (1)}(\mu)&=& 0.
\eea
This result
 coincides with the one obtained
previously in \cite{MJ1,EM},
but now, in addition, we have  shown explicitly, by means of
equations (\ref{eq025}), that the $a_{0}$, ${\hat a}_{1}$ and
${\hat a}_{8}$  coefficients are independent at one-loop
on the particular scheme used to renormalize both the fundamental and
the electroweak chiral lagrangian. Of course, they are still
dependent on the particular scheme
used to renormalize the coefficients itself. However there is
a combination of $a_{0}^{b}(\mu)$ and
${\hat a}_{1}^{b}(\mu)$ finite and scale independent
\beq \label{eq033}
g^{2} s^{2} {\hat a}_{1}^{b}(\mu) - {2 \over 9}
c^{2} a_{0}^{b}(\mu).
 \eeq
 This particular combination together with
${\hat a}_{8}^{b}(\mu)$ which is finite and scale independent
too, define an observable quantity called $L$ (see \cite{EM})
\beq  \label{eq034}
L=g^{2} s^{2} {\hat a}_{1}^{b}(\mu) + g^{2}
c^{2} {\hat a}_{8}^{b}(\mu) - {2 \over 9}
c^{2} a_{0}^{b}(\mu).
 \eeq
It is not difficult to get a physical insight on the meaning of this
quantity. If one writes down the Higgs contribution ($\rm{Log}$ plus
finite part) to the leptonic width ($\Gamma_{l}$) and
the effective Weinberg angle (${\bar s}^{2}$) as given by the
effective
field theory (in terms of the $a$'s) both quantities ($\Gamma_{l}$
and
${\bar s}^{2}$) will depend on the cutoff of this theory (in our case
the $M_{H}$). However one can  combine these two quantities to
construct
a $\rm{Log} M_{H}$ independent quantity which is precisely the one we
have called $L$.
Its value in the MSM is $L^{(1)}=0$. When the fundamental theory is
other than the MSM this quantity $L$ is constructed to cancel the
$\mu$ dependence (see \cite{EH92}).

Going back to eqs. (\ref{eq032}) the usual choice to renormalize this
coefficients
 is the modified Minimal Subtraction Scheme. After performing
this subtraction the renormalized coefficients at one-loop become
\bea \label{eq035}
a_{0}^{r \,(1)}(\mu)&=& {g^{\prime 2} \over 16 \pi^{2}}{3 \over 8}
\left(  -{\rm Log} {M_{H}^{2} \over \mu^{2}} + {5 \over 6} \right),
\nn \\
{\hat a}_{1}^{r \, (1)}(\mu)&=& {1 \over 16 \pi^{2}}
{1 \over 12} \left(
-{\rm Log} {M_{H}^{2} \over \mu^{2}} + {5 \over 6} \right), \nn
\\
{\hat a}_{8}^{r \, (1)}(\mu)&=& 0.
\eea
Notice that no $M_{H}^{2}$ contribution enters into these
coefficients due to the screening theorem \cite{Vel} at one-loop.

\bigskip
\noindent{\large \bf 3.2 $1/M_{H}^{2}$ correction and next order
operators}
\bigskip

If we  go one step further and expand the bare self-energies
(\ref{eq030}) up to the $1/M_{H}^{2}$ correction using
eqs. (\ref{apa3} - \ref{apa4}) of Appendix B
 one would get
\bea  \label{eq136}
 {\Sigma_{0 \, (H) \{ MSM \}}^{WW (1/M_{H}^{2})}}
(p^{2})&=&
-{g^2 \over 16 \pi^{2}} {1 \over M_{H}^{2}} \left\{ {3 \over
4}{M_{W}^{4}} {\rm Log}{M_{H}^{2} \over M_{W}^{2}} -{1 \over
3}{M_{W}^{2} p^{2}} -{1 \over 48} p^{4} \right\},
 \nn \\
 {\Sigma_{0 \, (H) \{ MSM \}}^{ZZ (1/M_{H}^{2})}}
(p^{2})&=&
-{g^2 \over 16 \pi^{2} c^{2}} {1 \over M_{H}^{2}} \left\{ {3 \over
4}{M_{Z}^{4}} {\rm Log}{M_{H}^{2} \over M_{Z}^{2}} -{1 \over
3}{M_{Z}^{2} p^{2}} -{1 \over 48} p^{4} \right\}, \nn \\
 {\Sigma_{0 \, (H) \{ MSM \}}^{\gamma \gamma (1/M_{H}^{2})}}
(p^{2})&=& 0, \nn \\
 {\Sigma_{0 \, (H) \{ MSM \}}^{\gamma Z (1/M_{H}^{2})}}
(p^{2})&=& 0.
\eea
These contributions will distribute in the following way:  the
$p^{0}$
parts will enter into the coefficient $a_{0}$, the $p^{2}$ terms into
${\hat a}_{1}$
and ${\hat a}_{8}$ and the $p^{4}$ pieces into the next order
operators.
The explicit Higgs contribution to $a_{0}$, ${\hat a}_{1}$ and
${\hat a}_{8}$
 can
be deduced by introducing eq. (\ref{eq136}) into the general
equations (\ref{eq025}). The results are
\bea  \label{eq337}
a_{0}^{r \, (1/M_{H}^{2})}(\mu)&=&- {g^{\prime 2} \over 16 \pi^{2}}{3
\over 8}{1 \over
s^2 c^2 }{M_{W}^{2} \over M_{H}^{2}} \left( {\rm Log} {M_{H}^{2}
\over M_{Z}^{2}} - c^{4} {\rm Log}{M_{H}^{2} \over M_{W}^{2}}
\right),
 \nn \\
{\hat a}_{1}^{r \, (1/M_{H}^{2})}(\mu)&=&
{1 \over 16 \pi^{2}} {1 \over 3} {M_{Z}^{2}  \over M_{H}^{2}},
\nn \\
{\hat a}_{8}^{r \, (1/M_{H}^{2})}(\mu)&=&{1
\over 16 \pi^{2}}{1 \over 3} {1 \over
M_{H}^{2}}(M_{Z}^{2}-M_{W}^{2}). \eea
Notice that the last coefficient
${\hat a}_{8}^{r \, (1/M_{H}^{2})}$
is nonzero for the first time,
so it gives  an isospin breaking term proportional to $1/M_{H}^{2}$.
As we will see in the next section, due to the zero contribution of
the two-loop $M_{H}^{2}$ correction to this operator,  the
$1/M_{H}^{2}$ term becomes
the most important source of breaking (together with the one coming
  from
the $p^{4}/M_{H}^{2}$ piece that enters the dimension six operators).
 With respect to the $a_{0}$ and
${\hat a}_{1}$ contribution it is clear that for a `light' Higgs
mass
and a large suppression factor of the $M_{H}^{2}$ two-loop
contribution
the $1/M_{H}^{2}$ piece can be the dominant correction to the leading
one-loop result. We will see that this is indeed the case for a large
Higgs mass range.

  From eq.(\ref{eq034}) and eq.(\ref{eq337}) one can obtain the
$1/M_{H}^{2}$ contribution to
the observable $L$  at this order
\beq
L^{(1/M_{H}^{2})}={1 \over 16 \pi^{2}}\left( {M_{W}^{2} \over
M_{H}^{2}}
\right) {g^{2} \over 3 c^{2}} \left\{ 1 - c^{4} + {1 \over 4} \left(
{\rm Log} {M_{H}^{2} \over M_{Z}^{2}} - c^{4} {\rm Log}{M_{H}^{2}
\over M_{W}^{2}} \right) \right\}.
\eeq

Up to this point we still have not said anything on how the $p^{4}$
contribution enters into the matching equations.
It is clear that the matching equations for this piece
will be of the form
\bea
\sum_{i} c^{i}_{WW} t_{i}&=\ns& {g^{2} \over 16 \pi^{2}}{1 \over
M_{H}^{2}}{1 \over 48}, \nn   \\
\sum_{i} c^{i}_{ZZ} t_{i}&=\ns& {g^{2} \over 16 \pi^{2} c^{2}}{1
\over M_{H}^{2}}{1 \over 48}, \nn   \\
\sum_{i} c^{i}_{\gamma \gamma} t_{i}&=\ns&0, \nn \\
\sum_{i} c^{i}_{\gamma Z} t_{i}&=\ns&0,
\eea
where the l.h.s. stands for the contribution of the possible
dimension six operators whose coefficients we have called $t_{i}$.
The $c^{i}_{X}$ coefficients
will depend only on $g,g^{\prime}$ and $v$.
No renormalization constants could enter at this order in momenta.

It is obvious from the previous equations that in the best situation
 four of the $t$
operators  can be fixed from the gauge self-energies. In general,
if they are more than four, one would
expect to be able to fix  just a combination of all
the possible coefficients of the $t$
operators that  contribute to the gauge self-energies.

Finally depending on the
 energy scale at which we are working (always inside the range
allowed by Chiral Perturbation Theory, which is related with the
$M_{H}$ in the MSM) the
 $t$ operators could become important.

\bigskip
\noindent{\Large \bf 4 Two-loop Matching Conditions}
\bigskip

At the two-loop order the situation becomes much subtler than at
one-loop,
 both from a
technical and a conceptual point of view.
In this section we will
discuss how to solve
 the problem of losing  gauge invariance and at the same time
taking into account only
gauge invariant operators in the matching conditions at two-loop
order.

First of all, it is obvious that
 equation (\ref{eq020}) still holds if we use the
nonlinear
representation for the Goldstone bosons  ($M_{2}$) in the MSM
(in fact this is true to all orders). This means that
we
will only be concern  with the diagrams
involving at least one Higgs. This simple observation can simplify
the work a lot in a complete two-loop computation.

Let's now consider the different contributions to the transverse part
of the two-point connected Green functions.
 At the two-loop order we will have One-Particle-Irreducible and
One-Particle-Reducible diagrams contributing.

A typical renormalized transverse One-Particle-Irreducible
self-energy at two-loops will be in the MSM of the form
\bea  \label{eq132}
&{\hat \Sigma}_{\{ MSM \} }(p^{2})=\ns&f_{(2)}(p^{2},M_{W,Z},M_{H}) +
 f_{(1)}(
p^{2},M_{W,Z},M_{H}) \delta Z_{\{ MSM \}}^{(1)} \nn \\
&&+ f_{(0)}(p^{2},M_{W,Z})
\delta Z_{\{ MSM \}}^{(2)},
\eea
where the first term on the r.h.s. $f_{(2)}$ stands for the
contribution
 coming from
two-loop diagrams,  the second represents one-loop
diagrams $f_{(1)}$ with insertions of one-loop counterterms
$\delta Z_{\{ MSM \} }^{(1)}$
and the last are tree level $f_{(0)}$ diagrams with two-loop
counterterm $\delta Z_{\{ MSM \} }^{(2)}$ insertions.

In the effective chiral lagrangian the corresponding self-energy
would read
 \bea  \label{eq133}
&{\hat \Sigma}_{\{ ECL \} }(p^{2})=\ns&g_{(2)}(p^{2},M_{W,Z}) +
 g_{(1)}(
p^{2},M_{W,Z}) \left[ \delta Z_{\{ ECL \}}^{ (1)} +  a_{i}^{(1)}
+ b_{j}^{(1)} \right]  \nn \\
&& + g_{(0)}(p^{2},M_{W,Z}) \left[
\delta Z_{\{ ECL \}}^{ (2)} +  a_{i}^{(2)} \right] ,
\eea
where the $g$ functions are the equivalent to the $f$'s in the chiral
lagrangian with one important distinction, the $g$ coefficients do
not bring any Higgs contribution.
 The whole $M_{H}$ contribution
is contained  in the $a$'s, the renormalization constants,
and, possibly, in the $b$'s.

With respect to the second type of
diagrams, the reducible ones, it is not difficult to see that none of
these diagrams would give a
new contribution to the matching conditions.
 There are mainly two types of such
diagrams as shown in Fig. 2. The diagrams of type (a), which are
just a product of two gauge
 self-energies, are
together with
 their one-loop renormalization constants
exactly balanced in the matching conditions
 with the contribution coming from the
chiral lagrangian (renormalization constants plus $a$'s). And the
diagrams of type (b)
 with a scalar particle (Higgs or Goldstone boson) running in
the middle of the two blobs,
 will always be proportional to $p^{\mu} p^{\nu}$, so they will
never give a contribution to the transverse self-energy.

In conclusion if we restrict ourselves to the {\bf transverse}
part of the two-point connected Green function is enough
to consider just One-Particle-Irreducible diagrams.
Let me
remind you again that this is no longer true for the longitudinal
parts.

 Even if we restrict to the transverse part one sees
immediately that in principal at two loops
 all unwelcome possible BRST operators ($b$) could enter.
For instance, due to
the contribution
 of Irreducible diagrams to the two-loop self-energy
as the one plotted in
Fig.3,
one could have self-energies of Goldstone bosons in one internal leg
as well as contributions from longitudinal self-energies of
 gauge bosons (in fact the longitudinal as well as the Goldstone
boson self-energy can be related by means of the Equivalence Theorem
\cite{Eq,ETR,ETCP}).
 That means in the language of Chiral Perturbation Theory
 that the $a_{11}$, $a_{12}$ and the possible
 BRST operators ($b$) associated with the
longitudinal as well
as the Goldstone boson self-energies
could contribute to the
transverse two-loop self-energy. Notice that at one-loop none of
these
operators could enter into the transverse gauge field self-energies.

We will show in the following how one can avoid these unphysical
operators
 and for the same prize simplify
substantially the matching equations at two loops. In the
subsection 4.2 we
will analyze the
particular case of the leading $M_{H}^{2}$ correction that thanks to
the
screening theorem at one-loop allows us to get the contribution to
the $a$'s at two-loop order in a rather easy  way.

\bigskip
\noindent{\large \bf 4.1 Formal solution: rewriting  the matching
equations at two loops}
\bigskip

In order to fix the ideas we will concentrate first on a particular
 One-Particle-Irreducible
 topology entering a general two-loop gauge self-energy
 like the one of Fig.3
where one already finds all the possible problems that could appear
in the general case (with all topologies).
 Depending on the particles that run inside the loop we can
distinguish typically between different type of contributions as
shown in
Fig.4. for the MSM and in Fig.5 for the electroweak chiral
lagrangian.
Of course, this set of diagrams is not intended to be complete but
just representative of the type of diagrams that will be relevant in
our discussion.
The diagrams (a),(b) and (c)  of Fig.4 would correspond to what
we have called in the one-loop case $\Sigma_{0 (H) \{ MSM \} }$ while
(a),(b) and (f) of Fig.5
would enter $\Sigma_{0 (H=a) \{ ECL \} }$ . Moreover the diagrams (d)
and (c) of
Fig.4 and 5 respectively would fit into the light part of the
bare self-energies, and according to the discussion in the previous
section they will cancel against each other
in the matching equations.

Notice that the diagrams (a) and (b) of Fig.5
could, in principal, include the unwelcome  $a_{11}$, $a_{12}$ and
all
 possible $b$-BRST operators as explained above.
 On the contrary it is clear that none
of these operators can
 contribute to the diagram (f) of the
chiral lagrangian Fig.5, simply because these type of operators
come from the introduction into the
lagrangian of a gauge fixing term, and by construction they can only
enter at tree level to the longitudinal but not to the transverse
self-energy.
Although it is quite reasonable to hope that at the end of the day
all such
unwanted $b$-operators entering the diagrams (a) and (b) of Fig.5
will cancel when adding all the contributions in the transverse
self-energy it could be quite involved  to proof such an hypothesis,
and assuming
it from the beginning could drive us to an incorrect determination of
the contribution to the gauge invariant operators.

Our method will consist mainly of rewriting the
 $b$-operators
 in terms of known objects, so in a way they will be
projected out from
the matching conditions and we will be left with just gauge invariant
operators. In such a way, we will be sure that we are not including
in the gauge invariant operators contributions that correspond to the
unknown BRST operators and moreover, as all these extra  BRST
operators
 should cancel in
any physical quantity, we will not lose any physical information by
not
 fixing
them with
such procedure.

In order to project out the extra BRST operators we
will use the matching conditions
at one-loop. For the topology we are looking at we will need
to use the Goldstone and  longitudinal gauge boson self-energies at
one-loop. It is clear, according to the discussion on section 2 that
if we want to impose these two matching equations at one-loop ( which
are now
BRST-invariant and not gauge invariant equations) one should include
together with the gauge invariant operators all
 possible BRST operators (exactly the same ones that contribute to
the two-loop  diagrams (a) and (b) of Fig.5). Symbolically the
one-loop
matching equation between the Goldstone self-energy in the MSM and
the
electroweak chiral lagrangian is drawn in Fig.6. (a similar equality
can be raised for the longitudinal gauge self-energy). Analytically
this one-loop matching equation for the Goldstone boson (and the one
of the longitudinal gauge boson) would read, following the same
notation as in eqs. (\ref{eq132}) and (\ref{eq133})
(but now at one-loop)
\bea  \label{eq036}
&
 g_{(0)}(
p^{2},M_{W,Z}) \left[ \delta Z_{\{ ECL \}}^{ (1)} +  a_{i}^{(1)}
+ b_{j}^{(1)} \right] =\ns&
{\tilde f}_{(1)}(M_{H},M_{W,Z}) +
 {\cal O}(1/M_{H}^{2}) + \nn \\
&& f_{(0)}(p^{2},M_{W,Z}) \delta {Z}_{MSM}^{(1)}
 + {\cal O}(\epsilon^{2}),
\eea
where ${\tilde f}_{(1)}(M_{H},M_{W,Z})$ stands for the leading
one-loop
Higgs mass  contribution in the MSM to the Goldstone/longitudinal
gauge boson
self-energy coming from diagrams including at least one Higgs running
inside the loop (as diagram (c) of Fig.6 for the Goldstone boson
self-energy).

Diagrammatically if we now substitute not only the $b$'s but the
$a$'s and renormalization constants of the diagrams
 (a)+(d) and (b)+(e) of Fig.5 using the one-loop matching
conditions
for the Goldstone boson and longitudinal
gauge self-energy (that are of the form of
eq.(\ref{eq036})) the diagrams of the electroweak chiral lagrangian
of Fig.5 turn out into Fig.7 (diagrams (a)+(d) of Fig.5 goes into
(a)+(d)  of Fig.7 and (b)+(e) of Fig.5 into (b)+(e) of Fig.7).
Where the box means that the diagram should be calculated in two
steps.
First one should compute the self-energy inside the box making the
explicit expansion in
 $M_{H}$ and $\epsilon$ up to the order given by eq.(\ref{eq036}),
and afterwards evaluate the rest of the diagram.
This is the reason why we need to know eq.(\ref{eq036}) to this order
in $M_{H}$ and $\epsilon$, not to lose any $M_{H}$ or finite piece.
By a direct comparison between Figs.4 and 7 it is clear
what we have gained.
On the one hand, the $b$'s have disappeared and, on the other,
new cancellations arise. The diagrams (d), (f) and (g) of Figs.4
cancel exactly against the diagrams (c), (d) and (e) of Fig.7.
So at the end  one only needs to calculate the type of
diagram (a) and
(e) of Fig.4 and the difference between the diagrams (b)+(c) of
Fig.4 calculated as a whole against the same diagrams (a)+(b) of
Fig.7 in the two steps explained above.

Of course, this procedure must be done consistently and taking into
account all topologies at the same time not to overcount diagrams.
Exactly the same technique can be applied to all other topologies
 (see \cite{VV} for the topologies), for instance,
with a triangle inside the loop diagram instead of a self-energy.
Moreover it is clear that this is a recursive procedure that it can
be
automatically extended to higher loop orders. At each loop order we
will always be able to eliminate the $b$'s and simplify the matching
conditions by inserting the previous loop order equations.

This method has the additional advantage of
showing
us explicitly that when all possible topologies are considered and
after performing the substitution explained above for each topology
one can
 get rid  of all the one-loop
renormalization constants and $a_{i}^{(1)}$ coefficients of the
chiral lagrangian as well
as the corresponding one-loop renormalization constants of the MSM in
the matching conditions at two loops.
With one
important exception, those
 diagrams in the MSM with one-loop renormalization constants inserted
 into one-loop diagrams that has a Higgs running
inside the loop (like the diagram (e) of Fig.4) will survive. This
set of diagrams (that we will refer as $c-diagrams$) are precisely
the ones drawn in Fig.1 after inserting in them
one-loop renormalization constants
in all possible ways.

In other words we are now in position to make a precise statement on
the {\bf condition} that should be fulfilled in order that
$a^{(2)}_{0}$, $a^{(2)}_{1}+a^{(2)}_{13}$ and $a^{(2)}_{8}+
a^{(2)}_{13}$ be scheme independent quantities in the sense of being
independent on the particular scheme used to renormalize both
theories (as it happens at one-loop). The condition
for each coefficient
is the following:
 if
the total contribution of the $c-diagrams$ entering each
 gauge self-energy cancel
 when combined as in eq.(\ref{eq025}) (substituting the
 $\Sigma_{0 \, (H)}^{X}$ by the corresponding $c-diagram$
contribution)
then this particular coefficient is independent on the scheme
used to renormalize both theories.
This condition, moreover, can be easily extended to $\Delta \rho$,
$\Delta r$ and $\Delta \kappa$.
It is clear that this condition should be fulfilled by the
 two-loop $M_{H}^{2}$ contribution entering the $a$'s
due to the
direct relation with
observables of this contribution.

For the more general case of a complete two-loop computation this
condition should be checked.
However it is not our goal in this paper to check but just enunciate
this property.

\bigskip
\noindent{\large \bf 4.2 Leading $M_{H}^{2}$ contribution
 }
 \bigskip

If we are interested just in the leading $M_{H}^{2}$ correction
the scenario simplifies enormously.
 Due to the screening theorem we know that
 the $M_{H}^{2}$ contribution,
entering the one-loop diagrams with external gauge fields, is not
observable and it can be completely absorbed
in the renormalization constants of the MSM (order $\epsilon$
included). Then
it is evident from the matching conditions eq.(\ref{eq016}) that all
the coefficients of the chiral operators at
one-loop  together with the renormalization constants of the
electroweak chiral lagrangian
will have at most a logarithmic Higgs dependence (including the order
$\epsilon$). This point is particularly evident in the on-shell
scheme.
 Moreover it has already been
pointed out that the coefficients $g_{(1)}$ do not
depend on the Higgs mass. As a consequence the second term
on the r.h.s. of eq.(\ref{eq133})
cannot be of  order $M_{H}^{2}$ and we can throw it
  from the matching equations. Then  the r.h.s. of the matching
conditions (the electroweak chiral lagrangian part)
 will have at two loops, if we are just regarding the leading
$M_{H}^{2}$
contribution, exactly the same form as the  one-loop matching
equations substituting
\beq \delta Z_{ECL}^{(1)} \rightarrow \delta Z_{ECL}^{(2)} \qquad
\qquad a_{i}^{(1)} \rightarrow  a_{i}^{(2)}. \eeq

 If we want to use one of the existing
calculations in the literature of the two-loop $M_{H}^{2}$
contribution
to the self-energies, the complete computation of Veltman and Van der
Bij \cite{VV,V2},
it is necessary first to clarify some points.
First, in their work
 they give the renormalized self-energies and not
the bare ones, however thanks to the screening theorem and
according to the conclusions of the previous paragraph one
can still  use eqs. (\ref{eq025})
after making the following
substitution everywhere
\beq  \label{eq026}
       {\Sigma_{0}}^{X}_{(H)} \rightarrow {{\bar \Sigma}^{X}}_{(H)
},\eeq
where ${{\bar \Sigma}^{X}}_{(H)}$ stands for the sum of the two-loop
diagrams involving a Higgs plus the Higgs contribution coming from
the  renormalization constants (the ${{\bar \Sigma}^{X}}_{(H)}$
can be extracted from \cite{VV,V2}).

Second, their result
was calculated using the linear representation $M_{1}$ of the
MSM, whereas we are using the $M_{2}$ parametrization for the
scalars fields. However
it is not difficult to show  following the simplified work of
\cite{BCS}
that if one is interested just in the $M_{H}^{2}$ contribution
the result will be the same in both parametrizations.

Let us comment briefly on that point.
In \cite{BCS} it was shown that by just taking into account Higgs
and
 scalar
particles inside the loop and working in Landau gauge one can
recover the same
results for the observables as \cite{VV} working in t'Hooft-Feynman
gauge and taking into account all particles.
  From a direct inspection of the relevant diagrams in \cite{BCS}
one can see that there is just one internal vertex
 involving Higgs and
Goldstone bosons that could produce different Higgs contributions
depending on the used parametrization (linear or nonlinear). This
vertex is in the nonlinear parametrization
\beq \label{eqx37}
{2 \over v} (\partial_{\mu} \pi \partial^{\mu} \pi \rho).
\eeq
We will rewrite it in a more useful way
\beq \label{eq037}
{2 \over v} (\partial_{\mu} \pi \partial_{\mu} \pi \rho) =
-{2 \over v} \Box \pi \pi \rho
+ {1 \over v} \pi^{2} \Box \rho,
\eeq
where the first term of eq.(\ref{eq037})
simply cancels one Goldstone boson propagator if we are working in
Landau gauge.
And the second term of eq.(\ref{eq037}) tell us that in any vertex
$\pi \pi \rho$  we will always
find that the propagator of the Higgs  comes with a $p^{2}$ in the
numerator and we can split it up into two pieces
\beq \label{eq138}
{p^{2} \over p^{2} + M_{H}^{2}} = 1 - {M_{H}^{2} \over p^{2} +
M_{H}^{2}}. \eeq
The second piece on the r.h.s.  always generates  an explicit
$M_{H}^{2}$.
This piece is exactly the same we find in the linear representation
of the
MSM. Then according to  \cite{BCS} the leading diagrams are those
with an explicit
Higgs mass, so
all the other pieces (the first on the r.h.s of eq.(\ref{eq037})
 and the first of eq.(\ref{eq138})
will always give subleading contributions). Of course, the
equivalence
between both parametrizations can be proved easily just to the
leading $M_{H}^{2}$ contribution if, instead, one was interested
 in performing the complete two-loop
calculation (and survive!) one should better choose directly
parametrization $M_{2}$,
because, as it has already been pointed out, only in this
parametrization
the light part of the self-energies of both theories cancel exactly
in
the matching equations. In any other parametrization one would need
to
calculate lots of two-loop diagrams involving only light particles
that contribute differently in the MSM than in the effective chiral
lagrangian.

If we now introduce the ${\bar \Sigma}^{X}_{(H)}$
 into the expressions for the
$a$'s in terms of the self-energies eqs.(\ref{eq025}) having
performed
the substitution (\ref{eq026}). We will end up with the following
two-loop Higgs contribution to the $a$'s
\bea \label{eq039}
a_{0}^{r \, (2)}&=&
{\left({1 \over 16 \pi^{2}} \right) }^{2}
{g^{2} g^{\prime 2} \over 64} \left( {M_H^{2} \over M_{W}^{2}}
\right) \left( 3 \pi^{2} + 9 \pi \sqrt{3} -108 C - {21 \over 2}
\right) , \nn \\ {\hat a}_{1}^{r \, (2)}&=&
{\left({1 \over 16 \pi^{2}} \right) }^{2}
{g^{2}  \over 288} \left( {M_H^{2} \over M_{W}^{2}}
\right) \left( -{28 \over 3} \pi^{2} + 9 \pi \sqrt{3} + 35 \right) ,
\nn \\ {\hat a}_{8}^{r \, (2)}&=& 0 ,
\eea
where the constant $C$ has a numerical value of 0.58598.
The relative importance of these corrections in front of the
$1/M_{H}^{2}$ will be addressed in section 5.

 At two loops the contribution coming from
$M_{H}^{2}$ to the observable quantity $L$ is

\beq \label{eq041} L^{(2)}=
{\left({1 \over 16 \pi^{2}} \right) }^{2}
{g^{4} s^{2} \over 288} \left( {M_H^{2} \over M_{W}^{2}}
\right) \left( -{37 \over 3} \pi^{2} + 108 C + {91 \over 2} \right) ,
\eeq
where again the cutoff cancellation does not take place as in the
subleading one-loop case.

\bigskip
\noindent{\large \bf 4.3 An extra screening in
  the photon self-energy  in the on-shell scheme
 }
 \bigskip

We will show in this section a very nice application of the chiral
description of the MSM.
We will prove  an `extra' screening
of the Higgs mass in the renormalized
photon self-energy in the on-shell scheme to all orders and the
automatic cancellation of the unphysical $M_{H}^{4}$ pieces in the
other gauge self-energies in this scheme at two-loops.

One possible presentation of this screening could  be: ``It can be
shown that
 the renormalized photon
self-energy in the on-shell scheme  will be at most
of order $M_{H}^{2 (n-2)}$ at the nth-loop order,
 while all other renormalized gauge self-energies will be
at maximum of order $M_{H}^{2 (n-1)}$". In other words there is an
extra suppression factor of at least $1/M_{H}^{2}$ in the
renormalized  photon
self-energy with respect to
the other gauge self-energies to all orders.

In order to prove it we will make use of Chiral Perturbation
 Theory,
the screening theorem of Veltman \cite{Vel} and the properties of the
on-shell scheme.

Let us start by proving it at the two-loop order. The irreducible
part of the renormalized photon self-energy at two-loops
will be in general of the form given by (\ref{eq133}). However we can
be a little
bit more specific with respect to the last term in the r.h.s. of
(\ref{eq133}) and write
\bea  \label{eq042}
&{\hat \Sigma}^{\gamma \gamma (2)}_{\{ ECL \}
}(p^{2})=\ns&g_{(2)}(p^{2},M_{W,Z}) +
 g_{(1)}(
p^{2},M_{W,Z}) \left[ \delta Z_{\{ ECL \}}^{ (1)} +  a_{i}^{(1)}
+ b_{j}^{(1)} \right]  \nn \\
&&+ p^{2} s^{2} g^{2} (a_{8}^{(2)} - 2 a_{1}^{(2)}) + ( s^{2} \delta
Z_{W}^{(2)} + c^{2} \delta Z_{B}^{(2)} ) p^{2},
\eea
where the second term on the r.h.s. contains one-loop diagrams with
insertions of $a$ and $b$
operators together with one-loop renormalization constants.
 As it was explained in detail in the previous section due to
the screening theorem no $M_{H}^{2}$ contribution could enter this
term. This is, in fact, the reason why it is instrumental the use of
the electroweak chiral lagrangian instead of the MSM itself, where
the equivalent to this term (second in the r.h.s of eq.(\ref{eq132}))
has in general $M_{H}^{2}$ contributions
that will combine with those coming from the two-loop diagrams and
two-loop renormalization constants.
 Of
course, the first term on the r.h.s. of eq.(\ref{eq042}) that comes
  from two-loop diagrams involving only light fields
 cannot have any Higgs contribution too. At the end all what remains
in the electroweak chiral lagrangian
are the two-loop $M_{H}^{2}$ contributions entering  $\delta
Z_{ECL}^{(2)}$ and the $a_{i}^{(2)}$'s.
 To analyze
these remaining contributions we will take advantage of the fact that
in the on-shell
scheme we know how to write the renormalization constants in terms of
bare self-energies (\cite{Ho88}). In particular $\delta Z_{W}^{(2)}$
and $\delta Z_{B}^{(2)}$ will be
\bea   \label{eq043}
\delta Z_{W}^{(2)}&=& - {\partial {\Sigma_{0}^{\gamma \gamma (2)}}
\over \partial p^{2}}\vert_{p^{2}=0} - 2 {c \over s}
{\Sigma_{0}^{\gamma Z (2)} (0)
\over M_{Z}^{2} } + {c^{2} \over s^{2}} \left(
{\Sigma_{0}^{ZZ (2)}(M_{Z}^{2})
\over M_{Z}^{2}} - {\Sigma_{0}^{WW (2)}(M_{W}^{2}) \over M_{W}^{2}}
\right), \nn \\
\delta Z_{B}^{(2)}&=& - {\partial {\Sigma_{0}^{\gamma \gamma
(2)}} \over \partial p^{2}}\vert_{p^{2}=0} + {2 }{s \over c}
{\Sigma_{0}^{\gamma
Z (2)} (0) \over M_{Z}^{2} } - \left( {\Sigma_{0}^{ZZ (2)}(M_{Z}^{2})
\over M_{Z}^{2}} - {\Sigma_{0}^{WW (2)}(M_{W}^{2}) \over M_{W}^{2}}
\right), \eea
where again due to the screening theorem the relation between the
renormalization constants of the electroweak chiral lagrangian
and the bare self-energies are the same as the one-loop relations
 for what concerns the leading $M_{H}^{2}$
contribution. If we now use them together with eqs.(\ref{eq021}),
we obtain for
the two-loop renormalization constants
\bea  \label{eq044}
\delta Z_{W}^{(2)}&=&- {1 \over s^{2}}(s^{2}-c^{2}) g^{2}
{\hat a}_{8}^{(2)} + 2 g^{2}
 {\hat a}_{1}^{(2)} + 2 {c^{2} \over s^{2}} a_{0}^{(2)},
\nn \\
\delta Z_{B}^{(2)}&=& - g^{2} {\hat a}_{8}^{(2)} - 2
a_{0}^{(2)} - g^{\prime 2} a_{13}^{(2)}. \eea
Going back to eq.({\ref{eq042}) if we substitute the renormalization
constants
by their expressions in terms of the $a$'s one finds that the pure
two-loop
Higgs contribution cancel exactly and we are left with just the
subleading terms
\bea  \label{eq045}
{\hat \Sigma}^{ \gamma \gamma (2)}_{\{ ECL \}
}(p^{2})&=&g_{(2)}(p^{2},M_{W,Z}) +  {\tilde g}_{(1)}(
p^{2},M_{W,Z}) \left[  a_{i}^{(1)}
+ b_{j}^{(1)} \right] ,
\eea
where we have rewritten  the renormalization constants at one loop
in terms of $a$'s and $b$'s too. On the other hand, it is easy to see
that the  contribution to the photon self-energy
in the electroweak chiral lagrangian coming from the
reducible diagrams does not modify this conclusion. From Fig.2
the diagram (a)
could give at maximum a ${\rm Log}^{2} M_{H}^{2}$ contribution
 again due to the screening theorem, while
the diagram (b) does not contribute to the transverse part.
 Finally according to the
matching conditions ${\hat \Sigma}^{\gamma \gamma (2)}_{\{ ECL \}
}$ should be equal to ${\hat \Sigma}^{\gamma \gamma (2)}_{\{ MSM \}}$ up
to a certain
order in the $1/M_{H}^{2}$ expansion.
It means that the renormalized photon self-energy
in the MSM will combine all its $M_{H}^{2}$ terms including
insertions of one-loop renormalization constants
 in such a way that
they cancel in the on-shell scheme.
The theorem is then proved at two-loops.

Before looking to higher orders let us consider what  happens to the
rest of
gauge boson self-energies. If one substitutes the renormalization
constants by their
expressions in terms of the $a^{(2)}$'s one finds that
in the renormalized self-energies
all the $a$'s
($a_{0}$,
${\hat a}_{1}$ and ${\hat a}_{8}$) always appear
multiplied by  $p^{2}$. Now by using a
result of Einhorn and Wudka \cite{EW} based on a simple
power counting which tell us that at nth-loop order the vacuum
polarizations are proportional
at most to $M_{H}^{2 n}$, one can imply easily that the $a$'s, which
bring the whole $M_{H}$ dependence, can be at two-loops
 at most of order $M_{H}^{2 \times (2-1)}$.
It means that one has proven that in the on-shell scheme {\bf all}
the
renormalized self-energies can grow at most as $M_{H}^{2}$. Notice
that this is not longer true in other schemes, like the one
used by Van der Bij and Veltman,
where they found indeed a $M_{H}^{4}$ dependence that
at the end it will cancel in all observables.

Once again one can repeat the same reasoning at three
loops for the renormalized photon self-energy with the only
difference now
that the first subleading term that survives the cancellation will be
at maximum of order $M_{H}^{2}$
\beq \label{eqx1}
{\tilde g}_{(1)}(
p^{2},M_{W,Z}) \left[  a_{i}^{(2)}
+ b_{j}^{(2)} \right] .
\eeq
 Notice that it is  possible again to use
eqs.({\ref{eq043}) and eqs.(\ref{eq044})
(substituting the superindex (2) by (3)) because now we are
 interested only in the leading $M_{H}^{4}$ three-loop correction.

In order to go to higher orders one can take again the result of
Einhorn and Wudka \cite{EW} and imply immediately that the maximum
Higgs contribution entering the $a$'s will be at the nth-loop order
\beq  \label{eq046}
a_{i}^{(n)}  \sim M_{H}^{2 \times (n-1)},
\eeq
with $a_{i}^{(n)}$ being the nth-loop contribution to an $a_{i}$
coming from diagrams
including some Higgs particle running inside the loop.
 Then it is evident from eq.(\ref{eq045})
that at the nth-loop order the irreducible part of the renormalized
photon self-energy in the on-shell scheme can grow at most like
 \beq  \label{eq047}
{\hat \Sigma}^{(n) \gamma \gamma}_{\{ ECL \} } \sim  M_{H}^{2 \times
(n-2)}, \eeq
whereas the other three renormalized self-energies will grow at most
like
 \beq
{\hat \Sigma}^{(n) ZZ, WW, \gamma Z}_{\{ ECL \} } \sim  M_{H}^{2
\times (n-1)}. \eeq
For what concerns the nth-loop reducible diagrams, due to their
 splitting into a
string of irreducible subdiagrams, they  will always  be at minimum
one power of
$M_{H}^{2}$ suppressed with respect to the leading nth-loop
irreducible diagram.

Let me insist that this result depend absolutely on the scheme. For
instance, by looking to the result of Van der Bij and Veltman
 the renormalized photon self-energy
in the particular renormalization scheme defined by them \cite{VV}
has in
fact an $M_{H}^{2}$ dependence. The conclusion will be then twofold.
On the one hand, the on-shell scheme seems to be a much natural
scheme in comparison with other possibilities
because as we have seen it cancels automatically all spurious
$M_{H}$ dependence. And on the other, the combination between the
screening
theorem and the properties of the electroweak chiral lagrangian
provide
us with a useful tool to make computations sometimes more easily than
in the MSM itself.

\bigskip
\noindent{\Large \bf 5 Discussion on the hierarchy of
corrections
 }
\bigskip

Up to this point we have obtained the leading and the first two
subleading
contributions entering the  chiral operators that contribute to the
transverse gauge boson self-energies.
But we have not yet said anything concerning  the hierarchy of these
subleading corrections.

According to the results of the previous sections the Higgs
contribution  entering
the renormalized gauge boson self-energies in the on-shell scheme
 is of the following form
\bea
&{\hat {\Sigma}}^{X}_{(H)}(p^{2})=\ns&{1 \over 16 \pi^{2}} \left\{
 M^{2}_{W,Z}
+ M^{2}_{W,Z} {\rm Log} M_{H}^{2} +  p^{2} +  p^{2}
{\rm Log} M_{H}^{2} \right. \nn \\
&&+ \left.  {M_{W,Z}^{4} \over M_{H}^{2}}{\rm Log} M_{H}^{2} +  p^{2}
{M_{W,Z}^{2} \over M_{H}^{2}}
+  {p^{4} \over M_{H}^{2}} + {\cal O}\left( {1  \over M_{H}^{4}}
\right) \right\} \nn \\ &&+{1 \over {( 16 \pi^{2} )}^{2}} \left\{
M_{H}^{2} +
 {M_{H}^{2} \over M_{W,Z}^{2}} p^{2} +  {\cal
O}\left({\rm Log}  M_{H}^{2} \right) \right\}, \nn \\
&& + \ldots
\eea
where the terms inside the first braces are the one-loop
contribution,
the second braces would correspond to the two-loop, etc. Moreover
inside each  brace the Higgs contribution is organized as an
expansion in inverse powers of $M_{H}^{2}$.
 In the language
of Chiral Perturbation Theory all these contributions enter  into the
renormalization constants and the
coefficients of the operators of the derivative expansion
according to their order in momenta (for instance, all
the $p^{2}$ pieces will enter into the $a$'s or $b$'s
coefficients, the $p^{4}$ into the $t$'s and so on and so
forth).
 A typical $a$ coefficient will have the following
structure
\bea
&a_{i}=\ns&{1 \over 16 \pi^{2}} \left\{ c_{1} {\rm Log} M_{H}^{2} +
c_{2} + c_{3}{M_{W,Z}^{2}
 \over M_{H}^{2}} + c_{4} {M_{W,Z}^{2} \over M_{H}^{2}} {\rm Log}
M_{H}^{2} + {\cal O}\left( {1 \over M_{H}^{4}}\right) \right\} \nn \\
&& + {1 \over {( 16
\pi^{2} )}^{2}}
\left\{ d_{1}{M_{H}^{2} \over M_{W,Z}^{2}} + {\cal
O}({\rm Log} M_{H}^{2})\right\} + \ldots \eea

It is clear from the previous equation that depending on the Higgs
mass and
on the particular value of the coefficients $c$'s and $d$'s the
subleading contribution coming from the one-loop ($1 / M_{H}^{2}$)
could be more important than
the one coming from the two-loop result $M_{H}^{2}$. Of course,
one should  not forget when comparing these contributions that with
each new loop
order an extra
 $1 / 16 \pi^{2}$ factor appears.

In this last section we will put all the pieces together
to analyze the different contributions  to three particularly
interesting quantities
 $\Delta \rho_{b (H)}^{(se)}$, $\Delta r_{b (H)}^{(se)}$ and $\Delta
\kappa_{b (H)}^{(se)}$
 paying special attention to the hierarchy of corrections.

 \bigskip
\noindent{\large \bf 5.1 Leading and subleading Higgs contribution to
$\Delta \rho_{b (H)}^{(se)}$, $\Delta r_{b (H)}^{(se)}$ and $\Delta
\kappa_{b (H)}^{(se)}$
 }
 \bigskip

There exist in the literature other parametrizations of the
non-decoupling effects of a Higgs at low energies.
 Two of them
are the set of $\epsilon_{1}$, $\epsilon_{2}$ and $\epsilon_{3}$
parameters \cite{AltBar}, or the set $S$, $T$ and
$U$ \cite{Pesk}.
Their relation with the $a$'s at one-loop remain still valid at
two-loops (only for the leading $M_{H}^{2}$ correction) due to the
absence of a
 $M_{H}^{2}$ contribution coming from
the second term on the r.h.s of
 eq.(\ref{eq133}). For instance, the contribution of the $a$'s
to the bosonic part of the $\epsilon_{i}$ parameters is
 \bea \label{eq040}
2 a_{0}^{r}(M_{Z}) &\rightarrow & \epsilon_{1}, \nn \\
-g^{2}{\hat a}_{8}^{r}(M_{Z}) &\rightarrow & \epsilon_{2}, \nn \\
-g^{2}{\hat a}_{1}^{r}(M_{Z}) &\rightarrow & \epsilon_{3}.
\eea
The previous relations are easily obtained by writing
the $\epsilon$'s in terms of  gauge self-energies \cite{AltBar} of
the electroweak chiral lagrangian. Where
one should include
 together with the $a^{b}_{i}(\mu)=a^{r}_{i}(\mu)+\delta
a_{i}$ all divergent and  ${\rm Log} (\mu / M_{Z})$'s pieces
entering the self-energies (in that
way the divergence and the $\mu$ dependence cancel automatically).
 Notice that the $t$ operators cannot enter
into the $\epsilon$'s.

One can choose any of the above parametrizations to write down
in terms of them
three process independent quantities that are sensitive to the
symmetry breaking sector, namely
 $\Delta \rho$, $\Delta r$ and $\Delta \kappa$ \cite{MS80,BH88}.

We will use, instead, the
 electroweak chiral lagrangian constructed up to now (including for
the
first time terms of order $1/M_{H}^{2}$ ). The self-energy
contribution
coming from Higgs corrections, divergences and ${\rm Log} (\mu /
M_{Z})$
 terms
 of the electroweak chiral lagrangian to the
 bosonic part of
 $\Delta \rho$, $\Delta r$ and $\Delta \kappa$ is given by
\bea  \label{eq51}
&\Delta \rho_{b (H)}^{(se)} =\ns& 2 a_{0}^{r}(M_{Z}),  \nn \\
&\Delta r_{b (H)}^{(se)} =\ns& - \left[ 2 g^{2}
{\hat a}_{1}^{r}(M_{Z}) + 2 {c^{2} \over
s^{2}}
 a_{0}^{r}(M_{Z})
 + {g^{2} \over s^{2}} (c^{2} - s^{2})
{\hat a}_{8}^{r}(M_{Z}) \right] \nn \\
&&+ {c^{2} \over s^{2}} \left[ (c^2 - s^2 ) \sum_{i} c^{i}_{WW} t_{i}
- \sum_{i} c^{i}_{ZZ} t_{i} \right] M_{Z}^{2}, \nn \\
&\Delta \kappa_{b (H)}^{(se)} =\ns& {c^{2} \over s^{2}} \left[ 2
a_{0}^{r}(M_{Z}) + g^{2}
{\hat a}_{8}^{r}(M_{Z})\right]  +
g^{2}{\hat a}_{1}^{r}(M_{Z}) \nn \\
&&+ {c^{2} \over s^{2}} \left[  \sum_{i} c^{i}_{ZZ} t_{i}
- c^{2} \sum_{i} c^{i}_{WW} t_{i} \right] M_{Z}^{2}
+{c \over s} \sum_{i} c^{i}_{\gamma Z} t_{i} M_{Z}^{2}.
\eea

Since we are interested only in the self-energy contributions
 no  vertex nor  box corrections are included (their contribution can
 be found,  for
instance, in \cite{MS80,BH88}).

If we now substitute in eqs. (\ref{eq51}) the different Higgs
contributions that we have found in the previous sections
$a_{i}^{(1)}+a_{i}^{(1/M_{H}^{2})}+
a_{i}^{(2)}$ together with the $t_{i}$'s, one ends up with the
following expressions
\bea \label{last}
&\Delta \rho_{b (H)}^{(se)}=\ns&+{1 \over 16 \pi^{2}}  {3 \over
4} g^{\prime 2}\left(
-{\rm Log} {M_{H}^{2}\over M_{Z}^{2}} + {5 \over 6} \right) \nn \\
& &- {1
\over 16 \pi^{2}}{3 \over 4}{
g^{\prime 2}
 \over s^{2} c^{2}} \left( M_{W}^{2} \over
M_H^{2}
\right) \left( {\rm Log} {M_H^2 \over M_Z^2 }
 -c^4 {\rm Log} {M_{H}^{2}
\over M_{W}^{2}} \right)  \nn \\
& &+{\left({1 \over 16 \pi^{2}}\right)}^{2}
{1 \over 32 }
 {g^{2} g^{\prime 2}}
 \left( {M_{H}^{2} \over M_{W}^{2}} \right) \left( 3 \pi^{2} +
9 \pi \sqrt{3} -108 C - {21 \over 2} \right), \nn \\
&\Delta r_{b (H)}^{(se)}=\ns&-{1 \over 16 \pi^{2}}{11 \over 12} g^{2}
\left(
-{\rm Log} {M_{H}^{2} \over M_{Z}^{2}} + {5 \over 6} \right) \nn \\
& & + {1 \over 16
\pi^{2}}{3 \over 4}{g^{2} \over s^{2}} \left( {M_{Z}^{2} \over
M_{H}^{2}} \right) \left(
  {\rm Log} {M_{H}^{2} \over
M_{Z}^{2}}
 -c^4 {\rm Log} {M_{H}^{2}
\over M_{W}^{2}}
- {17 \over 36} s^{2}(1 + 2 c^{2}) \right) \nn \\ & & - {\left({1
\over 16 \pi^{2}}\right)}^{2} {1 \over
8 } g^{4} \left( {M_{H}^{2} \over M_{W}^{2}} \right) \left( {25 \over
108} \pi^{2} + {11 \over 4} \pi \sqrt{3} -27 C - {49 \over 72}
\right) , \nn \\
&\Delta \kappa_{b (H)}^{(se)}=\ns&
+{1 \over 16 \pi^{2}}{5 \over 6} g^{2} \left( -{\rm Log}
{M_{H}^{2} \over M_{Z}^{2}} + {5 \over 6} \right) \nn \\
& & - {1 \over 16
\pi^{2}}{3 \over 4}{g^{2} \over s^{2}} \left( {M_{Z}^{2} \over
M_{H}^{2}} \right) \left(
  {\rm Log} {M_{H}^{2} \over
M_{Z}^{2}}
 -c^4 {\rm Log} {M_{H}^{2}
\over M_{W}^{2}}
- {17 \over 36} s^{2}(1 +  c^{2}) \right) \nn \\ & & + {\left({1
\over 16 \pi^{2}}\right)}^{2} {1 \over
8 } g^{4} \left( {M_{H}^{2} \over M_{W}^{2}} \right) \left( {53 \over
108} \pi^{2} + {5 \over 2} \pi \sqrt{3} -27 C - {119 \over 72}
\right)
. \eea

It is quite illustrative to make a plot of these corrections as a
function of the Higgs mass. By looking at the diagram (b) of
Figs.8,
9 and 10 ($\Delta \rho_{b (H)}^{(se)}$, $\Delta r_{b (H)}^{(se)}$ and
$\Delta \kappa_{b (H)}^{(se)}$ respectively)
it is clear that the $1/M_{H}^{2}$ correction (solid line) is the
dominant over the $M_{H}^{2}$ coming from two-loops (dashed line) in
the whole region in between 200 Gev and 1 TeV.
We will consider the $1/M_{H}^{2}$ correction to be reliable over the
200 Gev region
to be able to throw terms of order $1/M_{H}^{4}$ safely
and maintain a gap large enough with the light particles.

In the (a) plots
the Higgs contribution given by eq.(\ref{last})
represented by a solid line
is compared against the leading one-loop part (dotted line), leading
one-loop plus two loops (dashed-dotted line) and one-loop leading
plus $1/M_{H}^{2}$ correction (dashed line) for a range of values of
the Higgs mass.
 The Higgs
correction coming from two-loops is only able to give a
contribution large  enough to be distinguished from the leading
one-loop line around 600 Gev for $\Delta \rho_{b (H)}^{(se)}$, 700
Gev for $\Delta \kappa_{b (H)}^{(se)}$ and 800 Gev for $\Delta r_{b
(H)}^{(se)}$. Moreover in all cases both subleading contributions
come
(as seen in the (b) diagrams) with opposite sign,  producing and
extra
cancellation at around the TeV region, in such a way that the
complete
result coincides with the leading one-loop. Only in the region in
between
280 and 400 GeV there is a deviation of the order of 20 $\%$ between
the complete result and  the leading one-loop due to the
$1/M_{H}^{2}$
correction to $\Delta \rho_{b (H)}^{(se)}$. This percentage reduces
to 15 $\%$ for $\Delta k_{b (H)}^{(se)}$ and 11 $\%$ for $\Delta r_{b
(H)}^{(se)}$. Of course, if
 the Higgs mass turns out to be over the TeV the two-loop $M_{H}^{2}$
contribution will dominate over the $1/M_{H}^{2}$, but then new
problems
concerning the reliability of the perturbative series could arise.

 \bigskip
\noindent{\Large \bf 6 Conclusions
 }
 \bigskip

We have extended in this paper
the one-loop effective chiral description of the MSM
by incorporating
into the coefficients relevant to the LEP1 physics
the first two subleading corrections, the inverse
Higgs mass contribution at the one-loop level and the leading
two-loop
 Higgs
contribution.

In order to disentangle the non-decoupling effects of the Higgs of
the
 MSM in
the low energy dynamics of the light fields we have implemented the
matching conditions.

The object that we have used to match both theories is the transverse
part of the connected Green functions.

Following a constructive technique and taking advantage of a
nonlinear
representation for the scalar fields in the MSM we have obtained
at one-loop a simple
expression for the coefficients $a_{0}$, ${\hat a}_{1}$ and
${\hat a}_{8}$
in terms of just the Higgs contribution to the MSM bare
self-energies. It is  manifest in these equations
the independence of the chiral coefficients on the scheme of
renormalization
chosen in both theories. One should have in mind that due to the link
established
in the matching conditions
between the renormalization constants of both theories once we fixed
the renormalization scheme in one theory it is fixed in the other.
 Moreover
we have demonstrated that the
 same equations obtained at one-loop for the chiral coefficients
can be used to get the
contribution to these coefficients coming from the
$1/M_{H}^{2}$ one-loop correction and, thanks to the screening
theorem, from the two-loop $M_{H}^{2}$ terms.

We have solved the problem of the apparent inconsistency of losing
gauge
invariance in the matching conditions at two-loops and at the same
time consider only  gauge invariant operators, by proposing a
novel formulation of the matching conditions at two-loops, easily
extensible
to all orders. This new formulation turns out to have very nice
properties.
It removes all the one-loop renormalization constants from the
matching
conditions at two-loops, except from those renormalization constants
inserted into one-loop diagrams with a Higgs running inside.

Finally we have shown two applications of the constructed electroweak
chiral lagrangian. On the one hand, we have proven an extra screening
of the Higgs mass into the renormalized photon self-energy together
with the
automatic cancellation of the unobservable $M_{H}^{4}$ terms in the
rest of gauge self-energies in the on-shell scheme. And on the other,
we have obtained the
self-energy contribution to the bosonic part of $\Delta \rho$,
$\Delta r$ and $\Delta \kappa$ pointing out the larger contribution
of the one-loop $1/M_{H}^{2}$ in front of the two-loop $M_{H}^{2}$ up
to the 1 TeV region.

\bigskip

\noindent { \Large \bf Acknowledgements}
\medskip

The author is indebted to D. Espriu, F. Feruglio and G. Degrassi for
a careful reading of the manuscript and many useful discussions. I
also
thank Paolo Ciafaloni for helpful discussions regarding the two-loop
computation in the MSM.
It is a pleasure to thank the theory group of the University of
Padova
for the kind hospitality and nice atmosphere when completing this
work. I acknowledge financial support from Ministerio de Educacion y
Ciencia (Spain).
\bigskip
\pagebreak

\pagebreak

\appendix
\noindent{\large \bf
Appendix A:
List of $SU(2)_{L} \times U(1)_{Y}$ Gauge Invariant
Operators } \\ %
\bigskip

The set of ${\it C}$ and ${\it P}$ and $SU(2)_{L} \times U(1)_{Y}$
gauge invariant operators ${\cal L}_{i}$ are \cite{Long}
\bea
%\tabline
%\mc{c}{} & & \mc{c}{}  \\
%\hline
&{\cal L}_{0}=&{1 \over 4} a_{0} v^2  {T}_{\mu} {T}^{\mu} \nn \\
&{\cal L}_{1}=&{1 \over 2} a_{1} g g' B_{\mu\nu} {\rm Tr}
 T W^{\mu \nu} \nn \\
&{\cal L}_{2}=&{i } a_{2} g' B_{\mu\nu} {\rm Tr}[{T} V^{\mu} V^{\nu}]
\nn \\
&{\cal L}_{3}=&-i a_{3} g {\rm Tr}[W^{\mu\nu}[V_{\mu},V_{\nu}]]
\nn \\
&{\cal L}_{4}=&a_{4} {\rm Tr}[ V_{\mu} V_{\nu}] {\rm Tr}[ V^{\mu}
V^{\nu} ] \nn \\
&{\cal L}_{5}=&a_{5} {\rm Tr}[ V^{\mu} V_{\mu}] {\rm Tr}[ V^{
\nu} V_{\nu}]\nn \\
&{\cal L}_{6}=&a_{6} {\rm Tr}[ V_{\mu} V_{\nu} ] T^{ \mu}
T^{\nu} \nn \\
&{\cal L}_{7}=&a_{7} {\rm Tr}[ V_{\mu} V^{\mu}] T^{ \nu}
{T}_{\nu} \nn \\
&{\cal L}_{8}=&-{1 \over 4} a_{8}{g^2 } {\rm Tr}[ T W_{\mu \nu} ]
{\rm Tr}[ T W^{\mu \nu} ] \nn \\
&{\cal L}_{9}=&-{i} a_{9} {g} {\rm Tr}[ T W^{ \mu \nu}]
{\rm Tr}[ T V^{\mu} V^{\nu} ] \nn \\
&{\cal L}_{10}=&a_{10} ({T}_{\mu} {T}_{\nu})^2
\eea
The list is completed with three more operators,
 ${\cal L}_{11}$ and ${\cal L}_{12}$ that
vanish when using the equations of motion over them and ${\cal
L}_{13}$
that can be absorbed in a redefinition of the previous eleven
operators \bea
&{\cal L}_{11}=&a_{11} {\rm Tr}[({\cal D}_{\mu} V^{\mu})^2] \nn \\
&{\cal L}_{12}=&a_{12} {\rm Tr}[T{\cal D}_{\mu} {\cal D}_{\nu}
V^{\nu}] {T} ^{\mu} \nn \\
&{\cal L}_{13}=&{1 \over 2} a_{13} ({\rm Tr}[T {\cal
D}_{\mu}V_{\nu}])^2
\eea
where
\bea
&&V_{\mu}= (D_{\mu}U)U^{\dag}\qquad  T=U \tau_{3} U^{\dag} \qquad
{T}_{\mu}= {\rm Tr} T V_{\mu}\qquad  \nn \\ \nn \\
&&{\cal D}_{\mu}
O(x)=\partial_{\mu} O(x) + i g \left[ W_{\mu},O(x) \right ] . \eea
\pagebreak

\appendix
\bigskip
\noindent{\large \bf Appendix B: Basic formulae for the $1/M_{H}^{2}$
expansions} \bigskip

In this appendix we will give the necessary formulae to calculate
 the large Higgs mass limit of the
one-loop diagrams containing a Higgs,
although in the paper only  the leading and the first subleading
term are taken.

The basic quantity in which we are interested in is
\beq \label{apa1}
\int_{0}^{1} dx x^{n} {\rm Log} D_{2}(M^{2}) ,
\eeq
where $D_{2}(M^{2})$ was already defined in eq.(\ref{eq31}). It is
clear by
looking at $D_{2}(M^{2})$ that in order to get a convergent expansion
in inverse powers of $1/M_{H}^{2}$ one should take as the expansion
term \cite{Muta}
\beq
f={-p^{2} x (1-x) \over M_{H}^{2} (1-x) + M^{2} x}
\eeq
Then eq.(\ref{apa1}) can be split up into two pieces
\beq \label{apa2}
\int_{0}^{1} dx x^{n} {\rm Log} D_{2}(M^{2})=I_{1}+I_{2} \eeq
where

\bea \label{apb3}
&I_{1}=\ns&\int_{0}^{1} dx x^{n}
{\rm Log} (M_{H}^{2}
 (1-x) + M^{2} x)  \nn \\
&I_{2}=\ns&\int_{0}^{1} dx x^{n} \sum_{k=1}^{\infty} {(-1)}^{k+1}
{f^{k} \over k} .
\eea

After a straightforward but quite tedious calculation one can get
 a closed expression for eq.(\ref{apa2}) as an expansion in inverse
powers of $1/M_{H}^{2}$ that will be particularly useful if one
wish to perform the calculation by using a computer.
In order to write it in a more practical way we will define
$I_{1}^{x}$ and $I_{2}^{x}$ as the expansions of the integrals
$I_{1}$ and $I_{2}$ up to ${\cal O}(1/(M_{H}^2)^{x+1})$
\bea \label{apb4}
I_{1}&=& {\rm Lim}_{x \rightarrow \infty} I_{1}^{x} \nn \\
I_{2}&=& {\rm Lim}_{x \rightarrow \infty} I_{2}^{x}
\eea
where

\bea \label{apa3}
&I_{1}^{x}=\ns&{1 \over n+1}\left\{{\rm Log} M_{H}^{2} +
\sum_{k=n+2}^{x+n+1}
\left( \begin{array}{c} k-1 \\ n \end{array} \right) {\left(
{ M^{2} \over M_{H}^{2} } \right)}^{k-n-1} {\rm Log}\left({ M^{2}
\over
 M_{H}^{2}} \right) \right. \nn \\
&& \left. - \sum_{k=0}^{n} {1 \over n+1-k}
\left[ 1 + \sum_{l=k+1}^{x+k}
\left( \begin{array}{c} l-1 \\ k-1 \end{array} \right) {\left(
{M^{2}\over M_{H}^{2}} \right)}^{l-k} \right] \right\}
\eea
and
\bea \label{apa4}
&\ns\ns I_{2}^{x}=\ns&-\sum_{k=1}^{x} {\left ( p^{2} \over M_{H}^{2}
\right )}^{k}
{1 \over
k (k-1)! } \left\{ \sum_{s=n+k+1}^{x+n+1} \left( \begin{array}{c}
s-1 \\ n+k \end{array} \right) {(s-n-2)! \over (s-n-k-1)!} {\left(
M^{2} \over M^{2}_{H} \right)}^{s-n-k-1} \right. \nn \\
\ns\ns&-\ns\ns &\ns \sum_{r=0}^{n+k-1}{1 \over (n+k-r)(n+k-r+1)}
\ns \sum_{s=r+1}^{x-k+r+1}
\left( \begin{array}{c} s-1 \\ r \end{array} \right) { (s-r+k-2)!
 \over (s-r-1)!}
{\left( {M^{2} \over M^{2}_{H}} \right)}^{s-r-1} \nn \\
\ns\ns&-\ns\ns &\ns\ns \left. \sum_{s=n+k+1}^{x+n+1} \left[ \left(
\begin{array}{c} s-1 \\ n+k
\end{array} \right) - \left( \begin{array}{c} s \\ n+k+1 \end{array}
\right) \right]
\left[ \sum_{l=0}^{k-2} {\left( \begin{array}{c} s-n-k+l-1 \\ l
\end{array} \right)} {(k-1)! \over (k-1-l)}
\right. \right. \nn \\
&+\ns\ns& \left. \left. {(s-n-2)! \over (s-n-k-1)!} {\rm Log}\left(
{M^{2}
 \over M_{H}^{2} }\right) \right]
{\left( { M^{2} \over M_{H}^{2}} \right)}^{s-n-k-1}
\right\}
\eea
where the summatory $\sum_{l=0}^{k-2}$ should be taken equal to zero
if $k-2 < 0 $.

\pagebreak
\noindent{\large \bf
FIGURE CAPTIONS:
 }

\bigskip
\bigskip
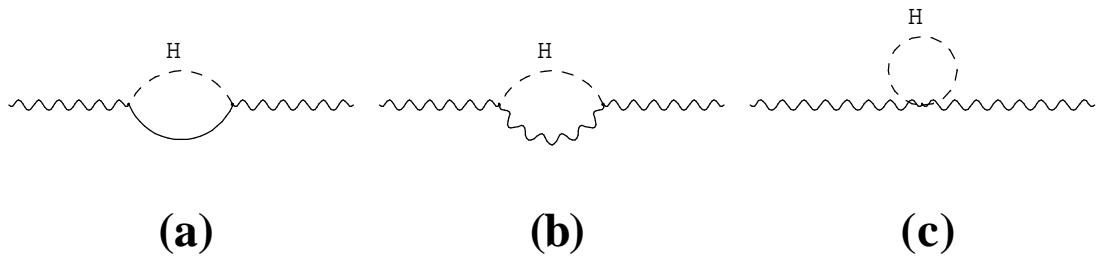
\noindent{{\bf Figure 1:}
 Diagrams contributing to ${\Sigma_{0 (H)\{MSM\}}}$
for the W and Z self-energy in the parametrization $M_{2}$.
The Goldstone (solid line) and gauge  boson particle (wavy line)
running
inside the diagrams (a) and (b) will be a $\pi^{+}$/$\pi^{3}$  and a
$W/Z$ particle
respectively depending on the gauge self-energy $W/Z$ we are
considering. Notice that the tadpole is exactly canceled by its
counterterm.}
\bigskip

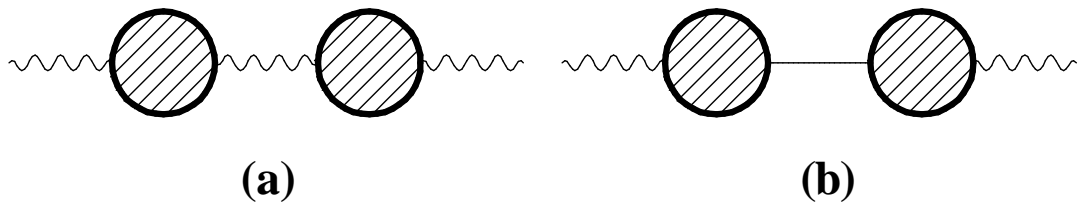
\noindent{{\bf Figure 2:}
One-Particle-Reducible diagrams entering a two-loop self-energy. The
shaded blobs stand for all possible one-loop diagrams. Figure (a)
represents those diagrams where the blobs are connected by a gauge
field, while in (b) the connection is done with a scalar
particle (Goldstone or Higgs).}
\bigskip

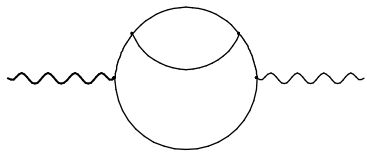
\noindent{{\bf Figure 3:}
 One of the Irreducible topologies entering a two-loop
gauge self-energy.}
\bigskip

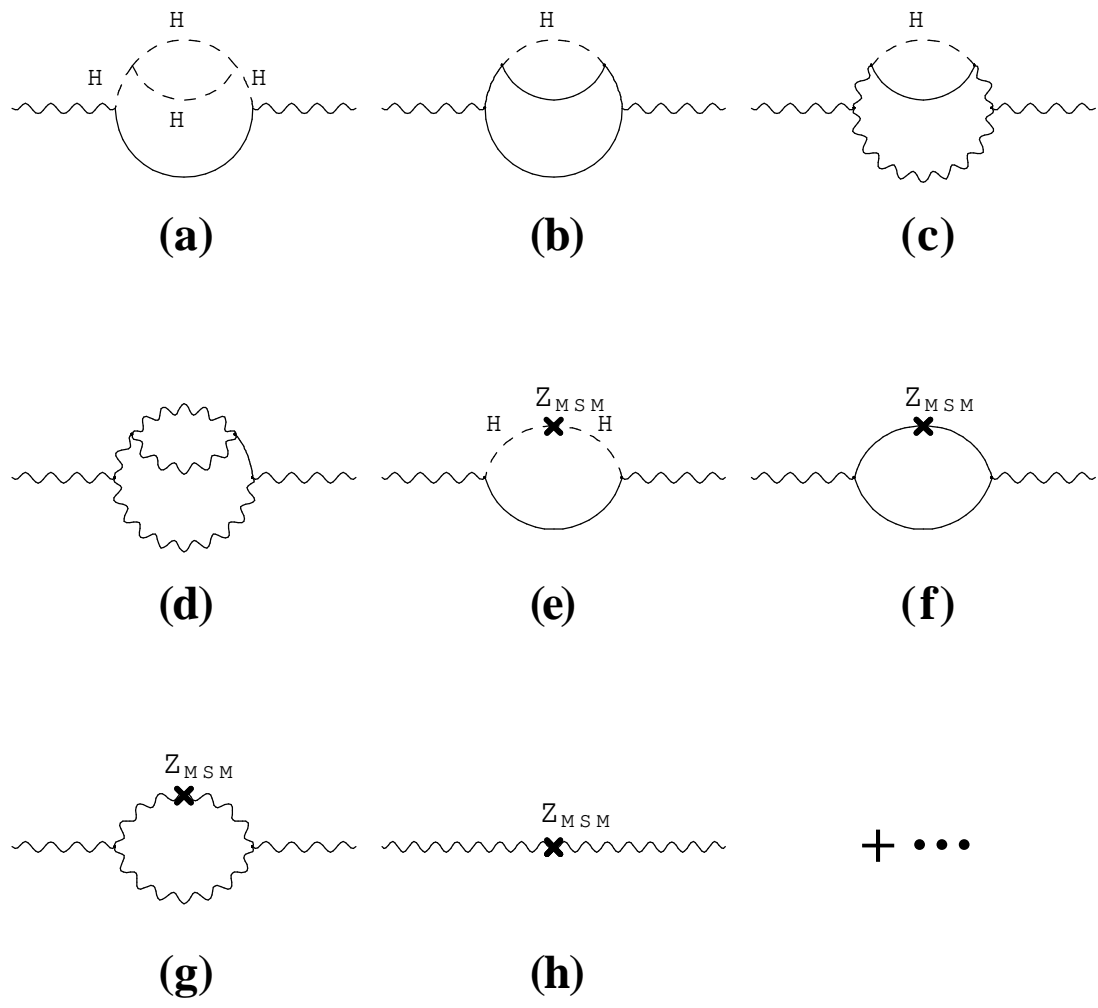
\noindent{{\bf Figure 4:}
 Set of diagrams representatives of
those entering
 the
topology of Fig.3 in the MSM.
The diagram (a) represents the class of diagrams with a Higgs in the
outer loop. The diagrams (b) and (c) stand for those diagrams with
only
light particles in the outer loop but a Higgs in the inner loop. The
diagram (d) contains only light particles. (e) (f) and (g) are
counterterm diagrams. Finally (h) is a
genuine two-loop counterterm. The dots stand for other
diagrams not involving Higgs (ghosts,...) that will
cancel automatically in the matching equations.
The number and position of all types of particles
can be changed  always according to the constraints given by the
representative of each class of diagrams. Goldstone and internal
gauge
boson lines are completely equivalent and can be interchanged freely,
that means that (b) and (c) are in fact equivalent but they are keep
separately to make the discussion in section 4 more clear. }
\bigskip

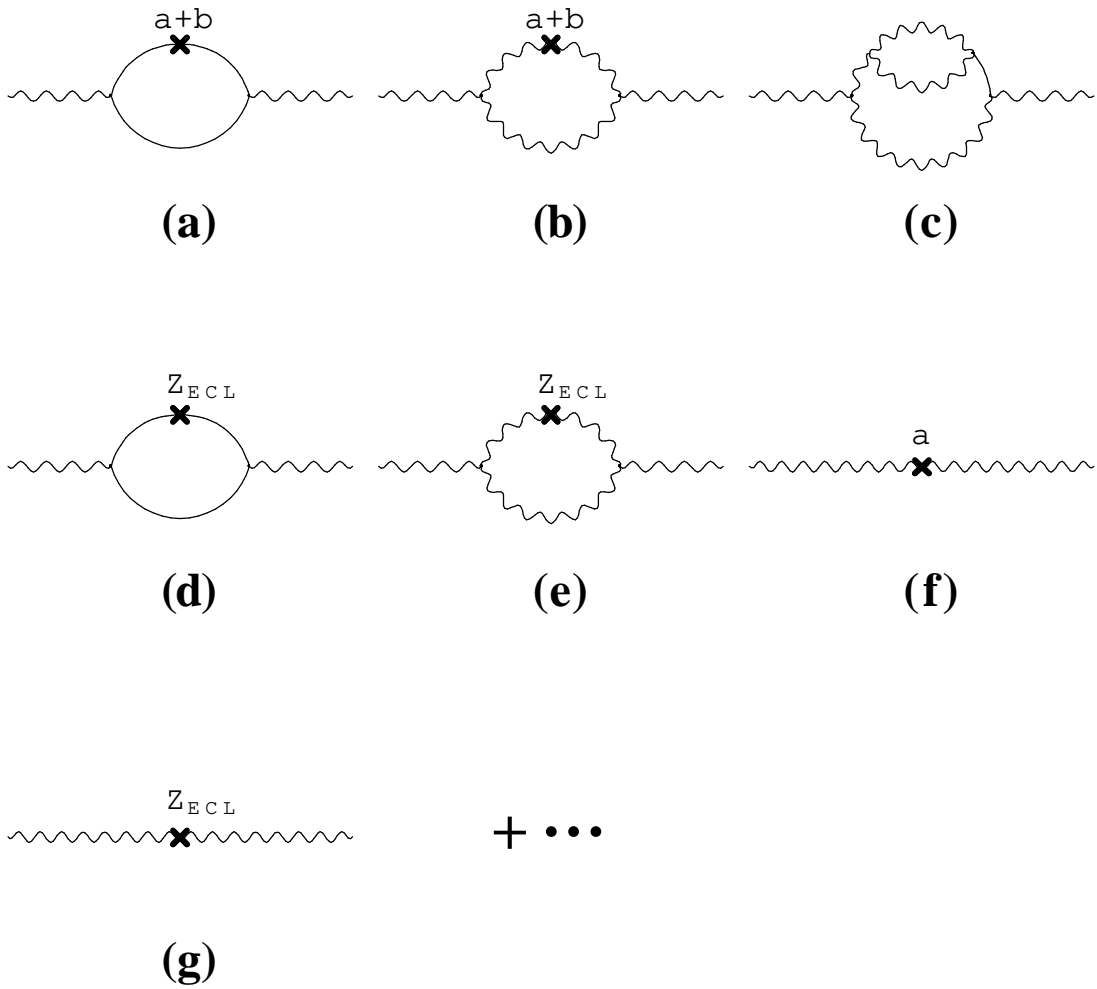
\noindent{{\bf Figure 5:}
 Set of diagrams representatives of
those entering
 the
topology of Fig.3 in the ECL. In the ECL no Higgs runs inside the
loops, its contribution is represented by the insertion of an $a$ or
$b$ operator.
The diagrams (a) and (b) represent the class of  diagrams with
scalars
and gauge particles and with an insertion of an $a$ or $b$ operator
at one-loop
order. Together with the renormalization constant insertion (d) and
(e) are represented by the second term in the r.h.s of
eq.(\ref{eq133}).
The diagram (c) does not have any Higgs contribution. Finally (f) and
(g) are the two-loop $a^{(2)}_{i}$ and counterterm
$Z_{ECL}^{(2)}$ insertions corresponding to the last term in
eq.(\ref{eq133}). The dots have the same meaning as in Fig.4
 }
\bigskip

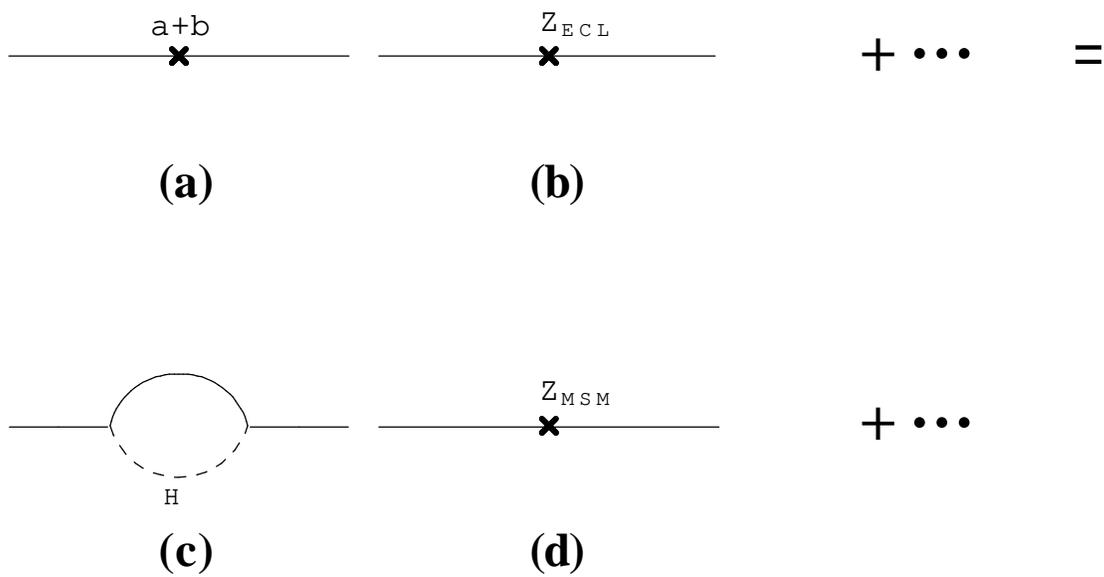
\noindent{{\bf Figure 6:}
  One-loop matching equation between the Goldstone boson
self-energy in the MSM and the ECL. Due to the gauge dependence of
this matching equation one should include, to be consistent, also $b$
operators. The dots in the l.h.s of the equality stand for one-loop
diagrams involving
light particles, whereas the dots in the r.h.s include in addition
other one-loop diagrams with a Higgs running inside that will be
relevant for other two-loop topologies
different from the one of Fig.3. A similar
equation can be raised for a longitudinal gauge field self-energy.}
\bigskip

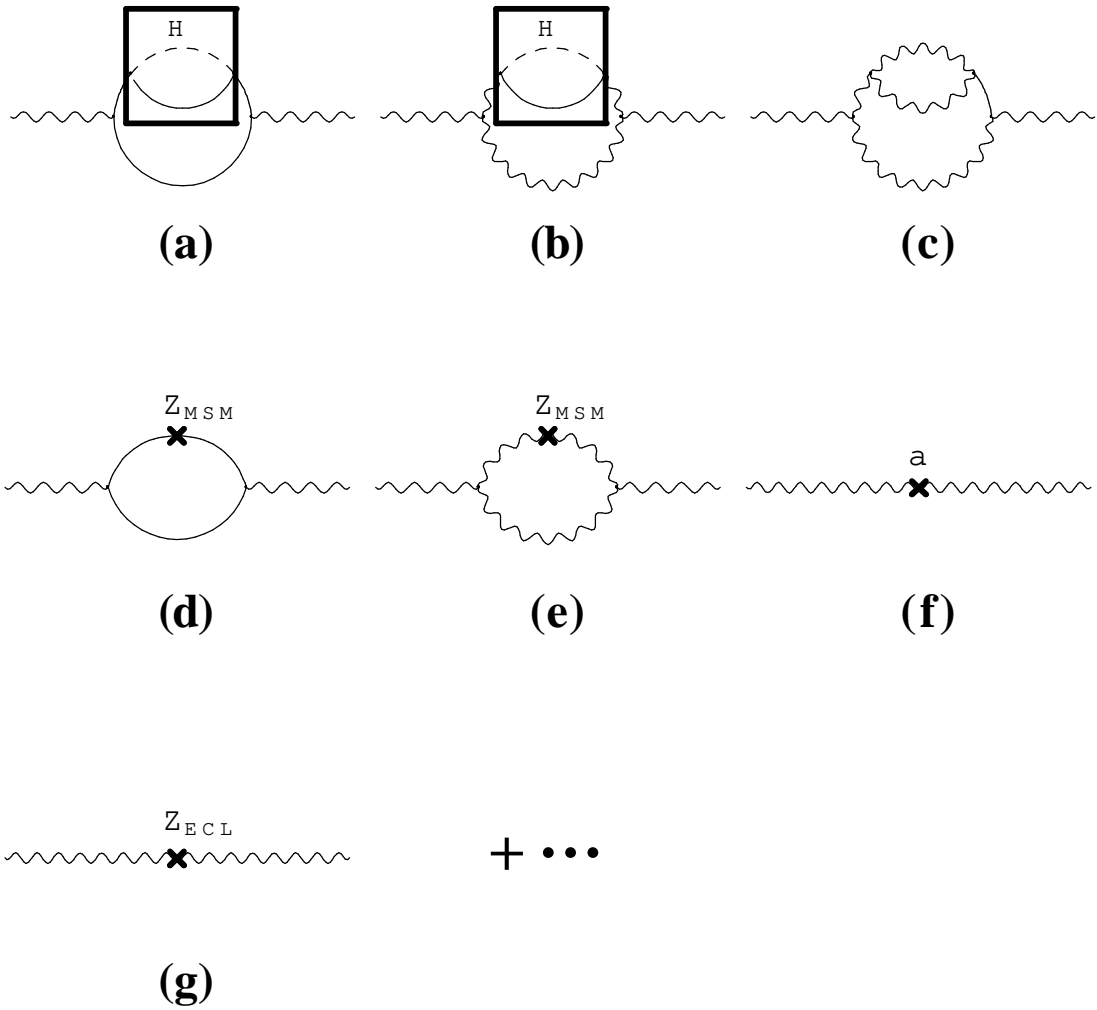
\noindent{{\bf Figure 7:} Set of diagrams transformed from those of
Fig.5 after substituting the $a^{(1)}$'s, $b^{(1)}$'s and
$Z_{ECL}^{(1)}$ of diagrams (a), (b), (d) and (e) of Fig.5
 by the $Z_{MSM}^{(1)}$
and the leading Higgs contribution coming from the diagram (c) of
Fig.6.
The box over the diagram (a) and (b) means that they should be
calculated in
two steps, first the large Higgs mass limit of the self-energy inside
the box and afterwards the rest of the diagram. Due to this simple
trick more cancellations arise in the matching equations.}
\bigskip

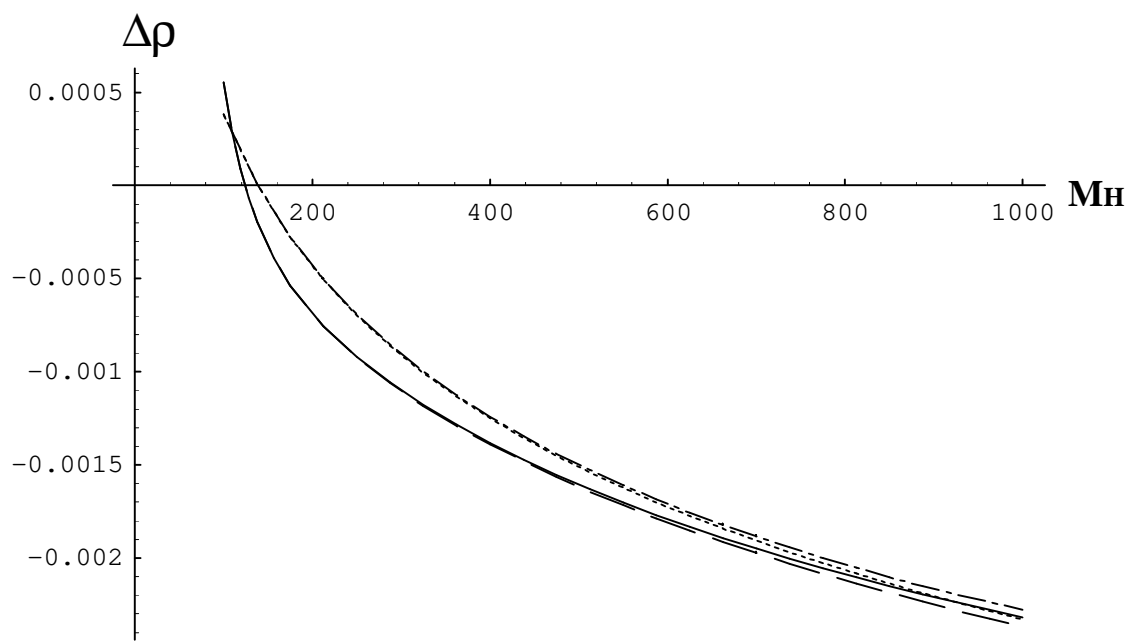
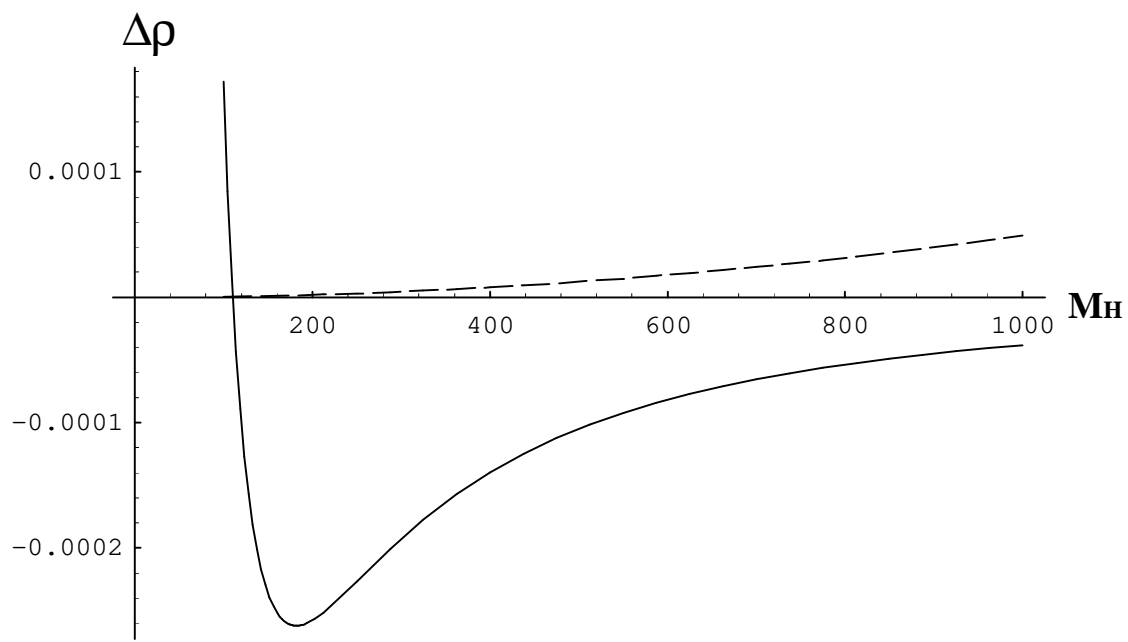
\noindent{{\bf Figure 8:}
 Self-energy Higgs contributions to the
bosonic part of $\Delta \rho$ plotted for a range of $M_{H}$
values up to 1 TeV.
In the diagram (a) the dotted line represents the
leading one-loop
Higgs contribution. The dashed line stands for the sum of the leading
one-loop plus the $1/M_{H}^{2}$ subleading. The dashed-dotted line is
the sum of the leading one-loop plus the two-loop $M_{H}^{2}$ piece.
Finally the solid line is the sum of all three contributions. In
the diagram (b) the two subleading contributions one-loop
$1/M_{H}^{2}$
(solid-line) and two-loop $M_{H}^{2}$ (dashed-line) are directly
compared.}
\bigskip

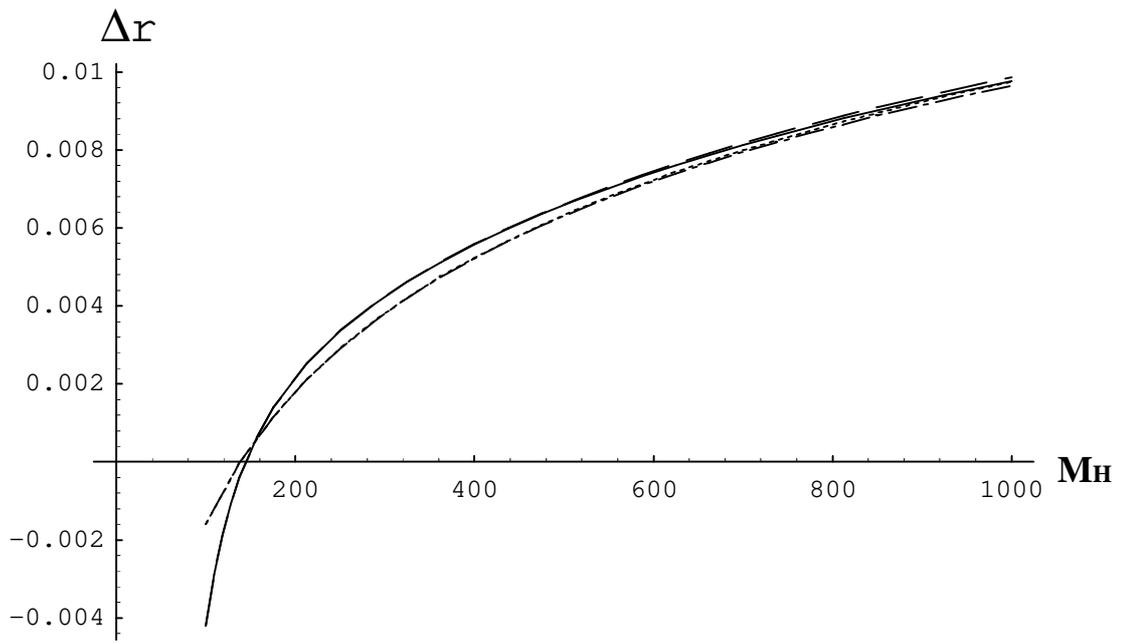
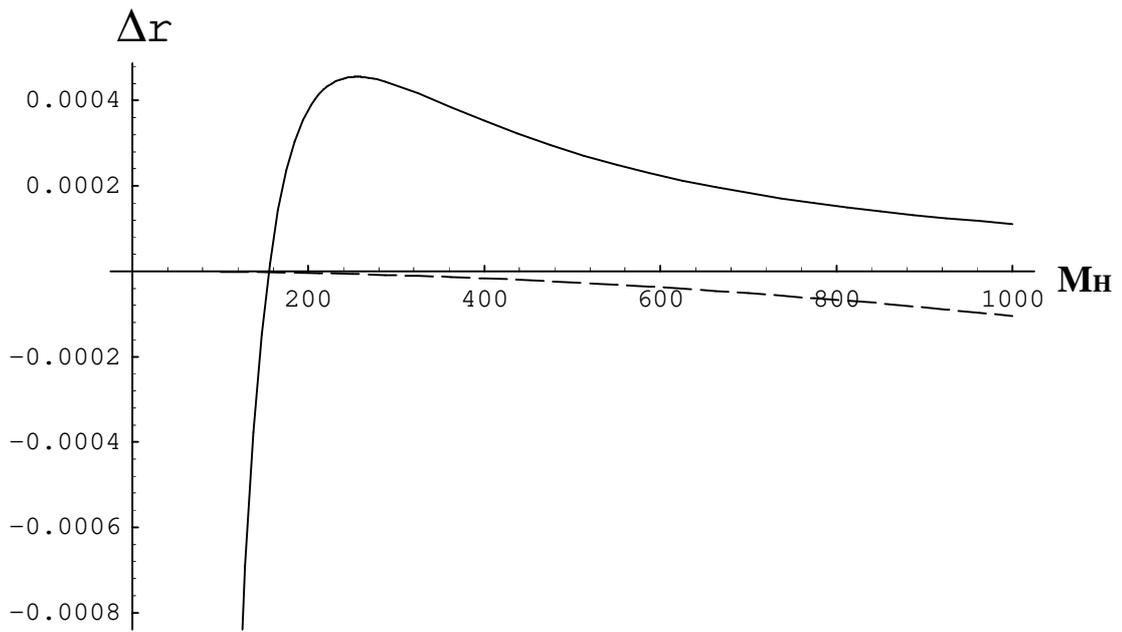
\noindent{{\bf Figure 9:}
 Self-energy Higgs contributions to the
bosonic part of $\Delta r$.
 Same conventions as in Fig.8.}
\bigskip

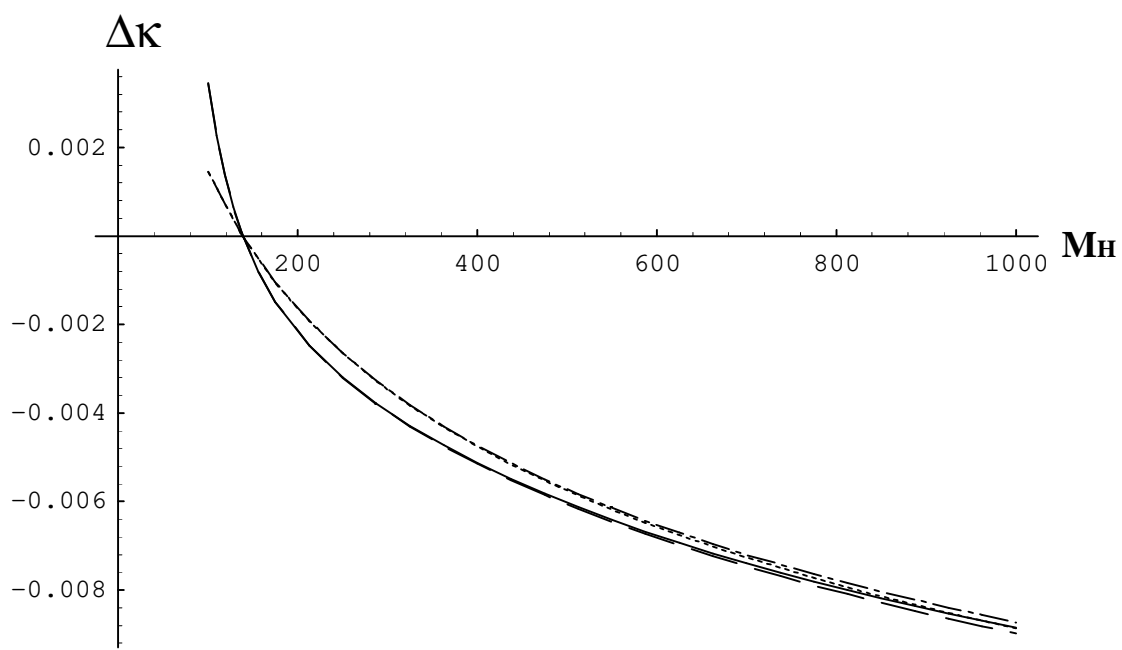
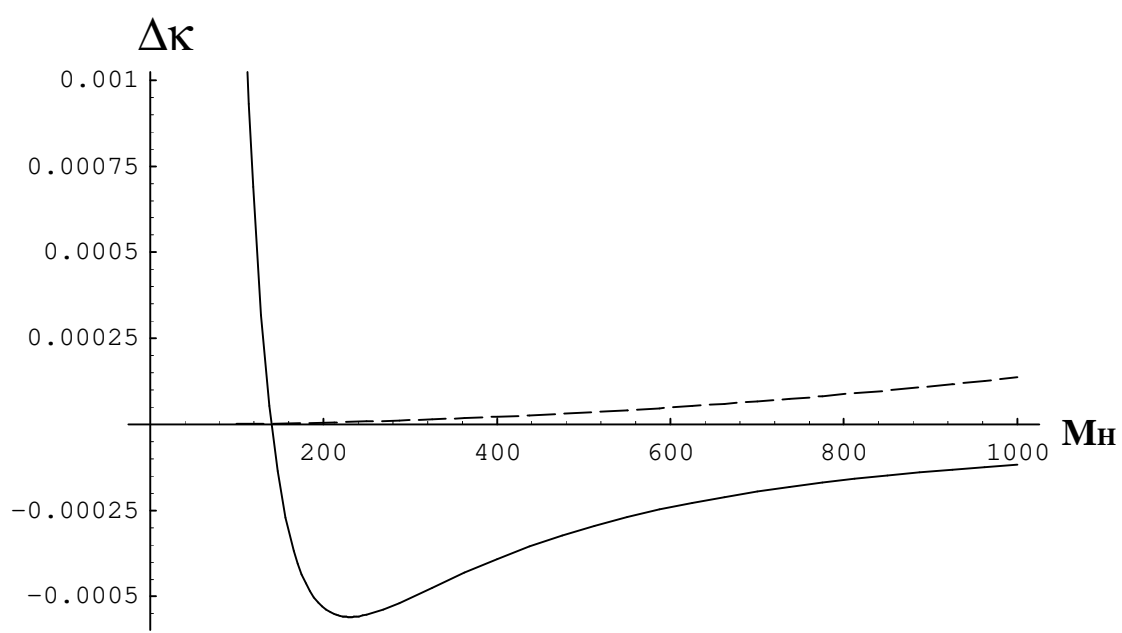
\noindent{{\bf Figure 10:}
 Self-energy Higgs contributions to the
bosonic part of $\Delta \kappa$.
 Same conventions as in Fig.8.}
\bigskip

\vfill

\bigskip

\end{document}